\documentclass[AMA,LATO1COL,doublespace]{WileyNJD-v2}
\usepackage{moreverb}

\newcommand\BibTeX{{\rmfamily B\kern-.05em \textsc{i\kern-.025em b}\kern-.08em
T\kern-.1667em\lower.7ex\hbox{E}\kern-.125emX}}

\articletype{Research Article}%
\received{11 October, 2017}
\revised{8 March, 2018}
\accepted{10 March, 2018}

\doiheadtext{DOI: 10.1002/fld.4504}

\usepackage{amssymb}                  
\usepackage[squaren,Gray]{SIunits}    
\usepackage{numprint}                 
\usepackage{caption}
\usepackage{algorithm,algorithmicx,algpseudocode}
\usepackage[capitalise]{cleveref}     

\begin{document}

\title{Resampling Strategies to Improve Surrogate Model-based  Uncertainty Quantification - Application to LES of LS89.}


\author[1]{Pamphile T. Roy*}
\authormark{Pamphile T. Roy \textsc{et al}}
\corres{*Pamphile T. Roy \email{roy@cerfacs.fr}}

\author[1]{Luis Miguel Segui}
\author[1]{Jean-Christophe Jouhaud}
\author[1]{Laurent Gicquel}

\address[1]{\orgname{CERFACS}, \orgaddress{42 Avenue Gaspard Coriolis, 31057 Toulouse cedex 1, \country{France}}}

\abstract[Abstract]{
Uncertainty Quantification (UQ) is receiving more and more attention for engineering applications in particular from robust optimization. Indeed, running a computer experiment only provides a limited knowledge in terms of uncertainty and variability of the input parameters. These experiments are often computationally expensive and surrogate models can be constructed to address this issue. The outcome of a UQ study is in this case directly correlated to the surrogate's quality. Thus, attention must be devoted to the Design of Experiments (DoE) to retrieve as much information as possible. This work presents two new strategies for parameter space resampling to improve a Gaussian Process surrogate model. These techniques indeed show an improvement of the predictive quality of the model with high dimensional analytical input functions. Finally, the methods are successfully applied to a turbine blade Large Eddy Simulation application: the aerothermal flow around the \textit{LS89} blade cascade.
}

\keywords{
Uncertainty Quantification ; Probabilistic methods ; POD ; Model reduction ; Aerodynamics ; Turbulent flow ; LES
}

\jnlcitation{\cname{%
\author{P.T. Roy}, 
\author{L.M. Segui},
\author{J.-C. Jouhaud}, and 
\author{L. Gicquel}} (\cyear{2018}), 
\ctitle{Resampling Strategies to Improve Surrogate Model-based  Uncertainty Quantification - Application to LES of LS89.}, \cjournal{International Journal for Numerical Methods in Fluids}, \cvol{2018;00:1-25}, DOI: 10.1002/fld.4504.}

\maketitle

\section{Introduction}
Design process and robust optimization are the major purposes of most engineering works dealing with Computational Fluid Dynamics (CFD), especially in aeronautical or automotive industry~\cite{duchaine2009}. Despite the large amount of work that has been devoted to the design of efficient optimization techniques, the design process still requires important investments (financial and human)~\cite{forrester2009}. As a consequence, design errors appear after the industrialization phase~\cite{draper1995} and the implications these can have may be critical. This partially explains why CFD tools have been used more and more in the past decades to decrease the number of iterations between conception and experiments to avoid irreversible errors during the preliminary design phase. 

Nowadays, CFD codes have reached maturity and represent more accurately physical flow phenomena. Complex flow simulations on high-resolution grids are possible thanks to the continuous developments in numerical models and in High Performance Computing (HPC). Nevertheless, deterministic simulations only provide limited knowledge about a system. Uncertainties in the numerical model as well as in the problem formulation or inputs are necessarily present and translate into uncertainties in the outputs~\cite{Sacks1989}.

In fact, the diversity of uncertainties on the CFD boundary conditions or initial conditions, as well as on model parameters (input data, geometry, simplification of the model physics, etc.) limits the validity of the simulations: the quantity of interest (QoI) can be easily affected and shadowed by the conjugation of all types of uncertainties. This assessment explains why Uncertainty Quantification (UQ) is now becoming a mandatory step in application-oriented modelling for operational and industrial purposes~\cite{degennaro2015,masquelet2017}. It provides insight into the level of uncertainty in the numerical simulation results but also gives access to the Sensitivity Analysis (SA) which aims at describing the respective influences of the input parameters on the QoI. The inclusion of UQ in a design optimization cycle hence allows manufacturers to design quicker and obtain better, cheaper and more robust (i.e. more stable) products.

Classical UQ methods, based on the Monte-Carlo approach, require a large number of CFD simulations~\cite{saltelli2007}, which quickly go beyond the limits of available computational resources (CPU cost). This is especially true when it comes to large dimensional problems, both with respect to the domain discretization and to the number of uncertain input parameters. The cost of the UQ study can however be significantly reduced when the CFD code is replaced by a surrogate model which is formulated in a parameter space and which is fast to evaluate at any set of uncertain variables~\cite{martin2005}. 

Two successful approaches for building a surrogate model are \textit{Polynomial Chaos}~\cite{najm2009} and \textit{Gaussian Process}~\cite{rasmussen2006}. The approach used in this work consists in combining Gaussian Processes with Proper Orthogonal Decomposition (POD) to build response surfaces through interpolation~\cite{braconnier2011,margheri2016}. In an industrial context---which is the case here---, some benefits of this method are:\textit{(i)} it does not require any prior knowledge on the probability distribution of the uncertainties on the input parameters ; \textit{(ii)} it does not need a specific sampling of the parameter space which could lead to \textit{curse-of-dimensionality} or mis-evaluation of the space ; \textit{(iii)} it provides an estimation of the predictive error ; and \textit{(iv)} the use of the POD reduces the number of predictors to compute. Details and comparisons can be found in~\cite{owen2015,roy2017}.

In any case, the number of CFD simulations that is required for the formulation of the surrogate model is defined by the complexity of the physics and the number of input parameters to take into account. This factor is paramount when considering costly numerical simulations.

The accuracy of an uncertainty quantification being directly correlated to the quality of the surrogate~\cite{iooss2010}, the present study aims at improving its construction by using two new strategies for resampling the parameter space. Industrial applications being targeted, a first UQ analysis of the \textit{LS89} case~\cite{arts1990} using Large Eddy Simulation (LES) is presented. LES are high-fidelity full 3D unsteady simulations. This approach comes at a high CPU cost which requires the use of High Performance Computing (HPC) resources.

The paper is tailored as follows; \cref{sec:method} starts by presenting the techniques employed to construct the surrogate model as well as its error assessment. Then the UQ tools used are detailed and \cref{sec:doe} describes the two proposed strategies. After this methodological presentation, \cref{sec:results} assesses the benefits of these strategies on several difficult analytical functions and presents the results on the application: the aerothermal analysis around the \textit{LS89} blade cascade. Finally, \cref{sec:ccl} will put a closure to this paper by summarizing its contributions along with potential directions for future works or applications.

\section{Surrogate model for UQ}
\label{sec:method}

In this study, the surrogate model is created using the \textit{BATMAN} (Bayesian Analysis Tool for Modeling and uncertAinty quaNtification) tool~\cite{roy2017} which is written in Python language. Using a non-intrusive approach, BATMAN allows to construct a surrogate model from any \textit{black-box} code. The result obtained is a POD-based Gaussian Process (pGP) surrogate which is computationally inexpensive and able to interpolate any new realization (or snapshot) $\mathbf{x_{*}}$ not in the sample composed of $N_s$ snapshots. BATMAN's workflow is detailed in \cref{alg:pGP}.

\begin{algorithm}
  \caption{POD-based Gaussian Process}
  \label{alg:pGP}
  \begin{algorithmic}[1]
  \State Choose the size of the sample $N_s$
  \State Pick $N_s$ samples in the input space $\mathbf{x}$ with a low discrepancy design of experiment (DoE)
  \State Perform a POD on the output
  \State Formulate the pGP surrogate $\mathcal{M}_{\text{gp}}$ on the POD's coefficients
  \State Resample the parameter space if necessary
  \end{algorithmic}
\end{algorithm}

\begin{figure}[h]
\centering
\includegraphics[width=0.8\linewidth,keepaspectratio]{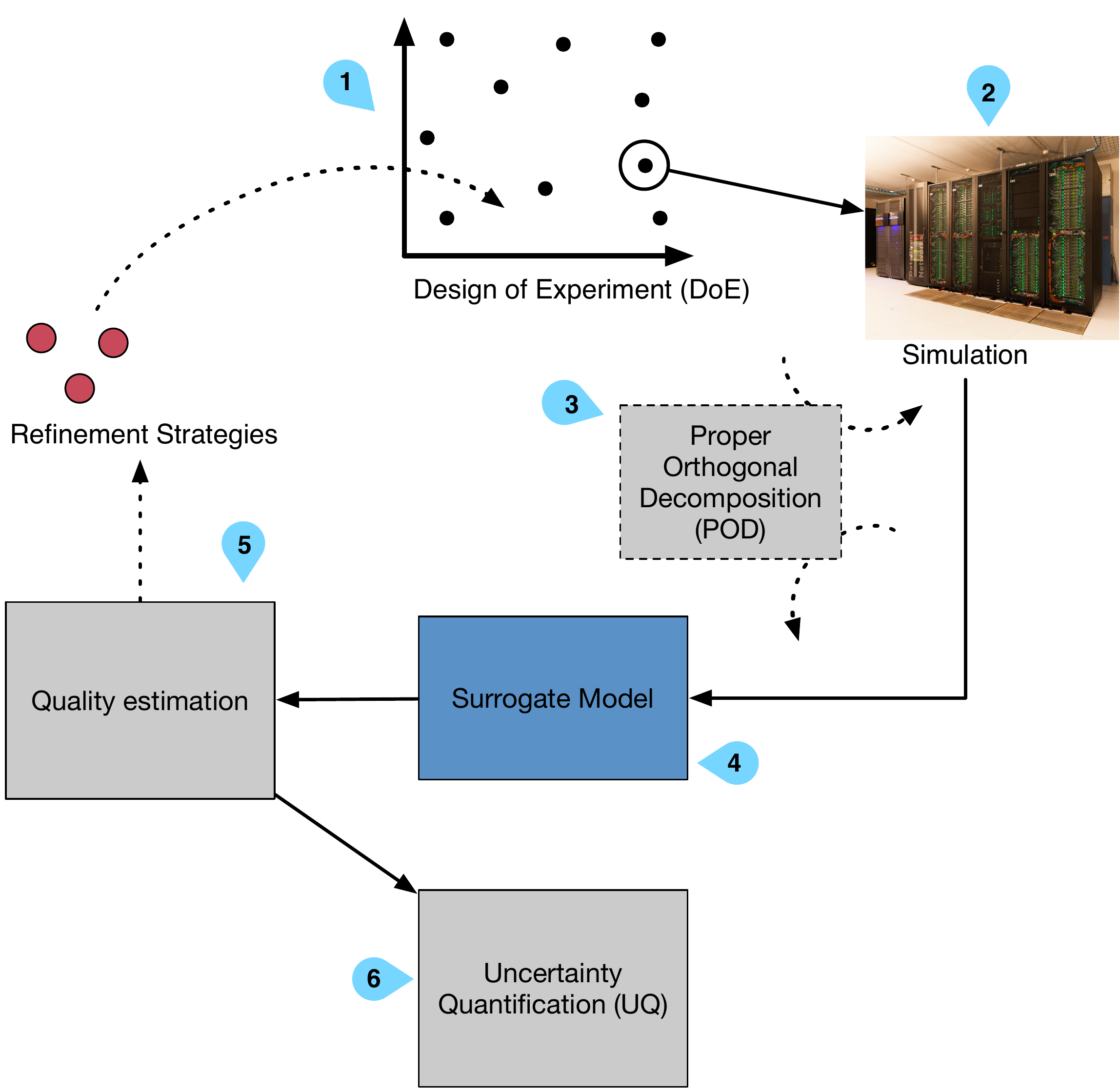}
\caption{Workflow to resample the parameter space in an Uncertainty Quantification framework.}
\label{fig:batman-chain}
\end{figure}

All different steps are detailed in the following sections. \Cref{sec:POD} presents the POD technique and \cref{sec:GP} reminds the Gaussian process technique. The quality of the surrogate is addressed in \cref{sec:error} and the UQ methods are detailed in \cref{sec:uq}. Finally, in order to improve the surrogate's quality, resampling strategies are presented in~\cref{sec:doe}. \Cref{fig:batman-chain} shows the workflow implemented in the BATMAN tool for constructing the surrogate model for UQ.

\subsection{POD}
\label{sec:POD}

A function $z$ can be approximate using a finite sum of terms:
\begin{align}
z(m,N_i) &\simeq \sum_{k} a_k(N_i) \phi_k(x),
\end{align}
with $m$ being the spatial discretization and $N_i$ a realization of the code, a snapshot. The functions $\phi_{k}(m)$ have an infinite representation and can be Fourier series, Chebyshev polynomials, etc. For a chosen basis of functions, a set of unique snapshot functions $a_k(N_i)$ arises. In the case of POD~\cite{chatterjee2000}, the basis functions are orthonormal which implies:
\begin{align}
(\phi_{k_1}, \phi_{k_2}) &= \left\{\begin{array}{rcl} 1 & \text{if} & k_1 = k_2   \\ 0 & \text{if} & k_1 \neq k_2\end{array}\right. ,\\
a_k (N_i) &= z(m,N_i) \cdot \phi_k(m),\nonumber
\end{align}

\noindent with $(\cdot, \cdot)$ the inner product. The principle of POD is to choose $\phi_k(m)$ such that the approximation of $z(m,N_i)$ is optimal in a least square sense. These orthonormal functions are called the \textit{proper orthogonal modes} of the function. Modes can be found using a \textit{Singular Value Decomposition} (SVD) of the snapshot matrix~\cite{sirovich1987}. Considering the snapshot matrix $\mathbf{A}$, gathering the output QoI computed spatially on $m$ for the $N_s$ snapshots:

\begin{align}
\mathbf{A} = \left[\mathbf{z}^{N_1}, \cdots, \mathbf{z}^{N_i}, \cdots, \mathbf{z}^{N_S} \right],
\end{align}

\noindent with $\mathbf{z}^{N_i}$ the \textit{i-th} snapshot vector stored as a column in $\mathbf{A}$. The SVD is a factorization operation of a matrix expressed as:
\begin{align}
\mathbf{A} = \mathbf{U} \mathbf{\Sigma} \mathbf{V}^T,
\end{align}

\noindent with $\mathbf{V} \in \mathbb{R}^{N_s \times N_s}$ diagonalizes $\mathbf{A}^T\mathbf{A}$, $\mathbf{U} \in \mathbb{R}^{m \times m}$ diagonalizes $\mathbf{A}\mathbf{A}^T$ and $\mathbf{\Sigma} \in \mathbb{R}^{m \times N_s}$ is the singular value matrix with its diagonal composed by the singular values $\sigma$ of $\mathbf{A}$. Thus, the initial matrix can be rewritten:
\begin{align}
\mathbf{A} = \sum_{i=1}^{r} \sigma_i u_i v_i^T,
\end{align}

\noindent with $u_i$ and $v_i$ the eigen vectors of respectively $\mathbf{U}$ and $\mathbf{V}$ which form an orthonormal basis. Also $r = \min (m, N_s)$ is the rank of the matrix. Due to the singular value matrix's null terms, the reduced form of the matrices is used which leads to $\mathbf{U} \in \mathbb{R}^{m \times N_{s}}$, $\mathbf{\Sigma} \in \mathbb{R}^{N_{s} \times N_{s}}$. Note that one can apply a filtering on the modes to only keep the basis vectors containing the highest energy of the system. For any $k<r$, an optimal approximation of rank $k$ of the snapshot matrix $\mathbf{A}_k = \mathbf{U}\mathbf{\Sigma}_{k} \mathbf{V}^T_{k}$ can be calculated by setting the $\sigma_{i>k} = 0$ in $\mathbf{\Sigma}$. These two steps allow to compress the data as only an extract of $\mathbf{U}$ and $\mathbf{\Sigma}$ need to be stored. But the real benefit is that a surrogate model does not need to be carried out for all points of the spatial discretization of the QoI but only for the matrix $\mathbf{\Sigma}_{k} \mathbf{V}^T_{k}$ (a column of this matrix corresponds to a snapshot).

\subsection{The Gaussian Process Regression}
\label{sec:GP}

A Gaussian Process (GP) is a collection of random variables which have a joint Gaussian distribution~\cite{rasmussen2006}. GP is equivalent to \textit{Kriging}~\cite{krige1989}. In this case, the random variable being the POD coefficients computed for each random input vector $\mathbf{x}$ of $N_s$: $f(\mathbf{x}) = (\mathbf{\Sigma}_{k} \mathbf{V}^T_{k})_{\mathbf{x}}$. A new prediction consists in a new column of $\mathbf{\Sigma}_{k} \mathbf{V}^T_{k}$. A \textit{Gaussian Process} $\mathcal{GP}$ is described by its mean $m(\mathbf{x})$ and covariance $k(\mathbf{x},\mathbf{x}')$---where $\mathbf{x}, \mathbf{x}'$ are different sets of inputs

\begin{align}
f(\mathbf{x})&\sim \mathcal{GP}(m(\mathbf{x}), k(\mathbf{x},\mathbf{x}')), \;\text{with} \\
m(\mathbf{x}) &= \mathbb{E}\left[ f(\mathbf{x})  \right], \nonumber \\
k(\mathbf{x},\mathbf{x}') &= \mathbb{E}\left[ (f(\mathbf{x}) -m(\mathbf{x}))(f(\mathbf{x}')-m(\mathbf{x}')) \right] \nonumber.
\end{align}

Here the covariance function $k$ (or kernel) is chosen as a squared exponential
\begin{equation} 
K = k(\mathbf{x}, \mathbf{x}') = \sqrt{\pi} \; \sigma_\mathbf{x}^2 \exp{- \frac{(\mathbf{x} - \mathbf{x}')^2}{2(\sqrt{2}l)^2}}, 
\end{equation}
where $l$ is a length scale that describes the trend in the data and $\sigma_\mathbf{x}$ is the variance of the output signal.
Then the GP model consists of a regression providing an interpolation $\hat{f}$ for a new set of input parameters $\mathbf{x_{*}}$:
\begin{align}
\hat{f}(\mathbf{x}_*)&=\bar{f}(\mathbf{x}_*) =  \sum_{i = 1}^{N_s}\alpha_i k (\mathbf{x}_i, \mathbf{x}_*), \;\text{with} \\
\mathbf{\alpha} &= (K + \sigma_n^2 I)^{-1}\mathbf{y}, \nonumber
\end{align}
where $\bar{f}$ is the mean realization, $\mathbf{x}_i$ the \textit{i-th} set of parameters, $\mathbf{y}$ the snapshot matrix and $\sigma_n$ is the variance of the input variables. Indeed, it is the mean realization of the conditioned process considering an artificial noisy observation which gives the prediction. The learning phase of the GP consists in selecting $l$, $\sigma_n$ and $\sigma_\mathbf{x}$ so that $f$ passes through or close to the dataset points. These hyperparameters are optimized using a differential evolution strategy. A key advantage of this predictor is that it provides an inference about its prediction variance
\begin{align}
\mathbb{V}[f(\mathbf{x}_*)] = k(\mathbf{x}_*, \mathbf{x}_*)-\mathbf{k}(\mathbf{x}_*)^T(K + \sigma_n^2 I)^{-1}\mathbf{k}(\mathbf{x}_*).
\end{align}

\subsection{Estimation of the error}
\label{sec:error}
To correctly adapt the number of snapshots to the required precision, the quality of the model has to be evaluated by comparing expected values and their estimations. A common indicator is to compute the predictivity coefficient $Q_2$~\cite{Marrel2009}: 
\begin{align} Q_2  = 1 - \frac{\displaystyle\sum_{i = 1}^{N_s} (f_i - \hat{f_i})^2}{\displaystyle\sum_{i = 1}^{N_s} ( f_i-\bar{f} )^2},\end{align}

\noindent with $\overline{f}$ denoting the mean value, $f_i$ the measured point and $\hat{f_i}$ its prediction by the model. When dealing with a non-analytical function the expected values are not known. However, there are two methods to evaluate the precision:

\begin{itemize}
\item  The sample can be divided into a validation set and training set. The model is built based on the training set and then evaluations are compared relative to the validation set. However, this technique requires that we do not use the validation simulations which is computationally costly when dealing with high fidelity numerical experiments.

\item 
Another approach is to estimate the quality by \textit{k-fold} Cross Validation~\cite{kohavi1995}. A particular case is the \textit{Leave-One-Out Cross Validation} (LOO) with~$k=n=1$---with $n$ the number of observations. The LOO technique derived from statistical learning theory requires the formulation of several surrogates. Each surrogate is built excluding one point from the evaluation sample; the accuracy of the surrogate is then calculated at this particular point. Removing point $p$ from $\hat{f_p}$ gives $\hat{f_p}^{(-p)}$ and leads to an error
\begin{align}
\epsilon_p &= \Vert f_p - \hat{f_p}^{(-p)} \Vert_2.
\end{align}
with the Euclidean 2-norm considered for $\mathbb{R}^m$. This is done all over the sample composed of $N_s$ snapshots to get the mean-square-error
\begin{align}
LOO &= \frac{1}{N_s} \sum_{p=1}^{N_s} \epsilon_p^2.
\end{align} 

Thus, an estimated predictivity coefficient $\hat{Q}_2$ can be retrieved
\begin{align}
\hat{Q}_2 = 1 - \frac{LOO}{\displaystyle\sum_{p = 1}^{N_s} \left( \hat{f}_p-\bar{\hat{f}} \right)^2} \quad .
\end{align}
\end{itemize}

As stated in~\cite{iooss2010}, this estimation tends to be pessimistic. Indeed, a snapshot is removed from an already very small sample which ultimately tends to lower the predictive quality of the model. However the indicator is stable---provided a correct sample size ($N_s > 10n_{dim}$, with $n_{dim}$ the number of dimension~\cite{forrester2007})---and as the quality increases, the difference between the estimation and the real quality vanishes.

\subsection{Uncertainty Quantification}
\label{sec:uq}

There are several methods to estimate the contribution of different parameters on quantities of interest~\cite{iooss2016}. Among them, sensitivity methods based on the analysis of the variance allow to obtain the contribution of the parameters on the QoI's variance~\cite{ferretti2016}. Here, classical \textit{Sobol'}~\cite{Sobol1993} method is used which gives not only a ranking but also quantifies the importance factor using the variance. This method only makes the hypothesis of the independence of the input variables. It uses a functional decomposition of the variance of the function to explore:

\begin{align}
\mathbb{V}(\mathcal{M}_{gp}) &= \sum_{i}^{p} \mathbb{V}_i (\mathcal{M}_{gp}) + \sum_{i<j}^{p}\mathbb{V}_{ij} + ... + \mathbb{V}_{1,2,...,p},\\
\mathbb{V}_i(\mathcal{M}_{gp}) &= \mathbb{\mathbb{V}}[\mathbb{E}(\mathcal{M}_{gp}|x_i)]\nonumber\\
\mathbb{V}_{ij} &= \mathbb{\mathbb{V}}[\mathbb{E}(\mathcal{M}_{gp}|x_i x_j)] - \mathbb{V}_i - \mathbb{V}_j,\nonumber
\end{align}

\noindent with $p$ the number of input parameters constituting $\mathbf{x}$. This way \textit{Sobol'} indices are expressed as
\begin{align}S_i = \frac{\mathbb{V}[\mathbb{E}(\mathcal{M}_{gp}|x_i)]}{\mathbb{V}[\mathcal{M}_{gp}]}\qquad S_{ij} = \frac{\mathbb{V}[\mathbb{E}(\mathcal{M}_{gp}|x_i x_j)] - \mathbb{V}_i - \mathbb{V}_j}{\mathbb{V}[\mathcal{M}_{gp}]}.\end{align}

\noindent $S_{i}$ corresponds to the first order term which apprises the contribution of the \textit{i-th} parameter, while $S_{ij}$ corresponds to the second order term which informs about the correlations between the \textit{i-th} and the \textit{j-th} parameters. These equations can be generalized to compute higher order terms. However, the computational effort to converge them is most often not at hand~\cite{iooss2010} and their analysis, interpretations, are not simple.

Total indices represents the global contribution of the parameters on the QoI and express as:
\begin{align}S_{T_i} = S_i + \sum_j S_{ij} + \sum_{j,k} S_{ijk} + ... \simeq 1 - S_{i}.\end{align}

For a functional output, as for the \textit{LS89} case---see~\cref{sec:ls89}---, \textit{Sobol'} indices can be computed all along the output and retrieve a map or create composite indices. As described by Marrel~\cite{marrel2015}, aggregated indices can also be computed as the mean of the indices weighted by the variance at each point or temporal step
\begin{align}
S_i = \frac{\displaystyle\sum_{l = 1}^{p} \mathbb{V} [\mathbf{f}_l] S_i^{l}}{\displaystyle\sum_{l = 1}^{p} \mathbb{V} [\mathbf{f}_l]}.
\end{align}

In this paper the indices are estimated using \textit{Martinez}' formulation. In~\cite{baudin2016}, they showed that this estimator is stable and provides asymptotic confidence intervals---approximated with Fisher's transformation---for first order and total order indices.

\section{Improving the Design of Experiment}

\subsection{Description of the new resampling methods}
\label{sec:doe}

Correctly sampling the parameter space is paramount as it is used to construct the model. Although the golden standard would be to perform a \textit{Monte Carlo} sampling, it would require an unreasonably large sampling which is unfeasible within a costly simulation environment or if considering \textit{real-time} applications. This constrains the number of simulations that can be performed. Cavazzuti~\cite{cavazzuti2013} provides a comprehensive description of the techniques used to generate the best DoE.

A good criterion for assessing the quality of a DoE technique is the discrepancy~\cite{iooss2010,fang2006}. It measures the uniformity of the points' coverage of the parameter space. Hence, \textit{low discrepancy sequences}, or \textit{quasi-random sequences}, have good filling properties of the space. To name a few, \textit{Sobol'} and \textit{Halton} sequences are known to perform well when dealing with low-dimensional spaces~\cite{damblin2013,saltelli2007}. Furthermore, an advantage over the traditional LHS~\cite{mckay1979} or optimized LHS~\cite{damblin2013} sampling is that the sample is deterministic. The sequence can be continued without losing any space-filling quality whereas with traditional LHS, the sample becomes suboptimal. Indeed, LHS algorithms require a number of points to create the sample.

This last observation motivates our choice of a low discrepancy sequence for sampling the parameter space. Indeed, this enables us to increase the design one simulation at the time. A complementary strategy consists in exploring the space using as few points as possible and then refine the exploration around zones of interest.

Starting from the work of~\cite{scheidt2006,braconnier2011} with the $\sigma$ method as a baseline, two novel strategies---LOO-$\sigma$ and LOO-\textit{Sobol'}---have been developed and are presented in this work. The common strategy is detailed in \cref{alg:refine}.

\begin{algorithm}
  \caption{Refinement strategy}
  \label{alg:refine}
  \begin{algorithmic}[1]
  \Require $N_{max}$, $threshold$
  \While{$LOO-quality < threshold$ and $N_s < N_{max}$}
    \State $\mathbf{x}_{L} \gets$ least stable point of the design
    \State $\mathcal{H_{L}} \gets$ maximum hypercube around $\mathbf{x}_{L}$
    \State $\mathbf{x}_o \gets \max \mathbb{V}[\mathcal{M}_{gp}]$, within $\mathcal{H_{L}}$
	\State Compute a new snapshot at $\mathbf{x}_o$
	\State Update pGP surrogate $\mathcal{M}_{gp}(\mathbf{x}_*)$
  \EndWhile
  \end{algorithmic}
\end{algorithm}

\begin{itemize}
\item Variance ($\sigma$), \hfill\\
As stated in \cref{sec:GP}, one of the main advantages of Gaussian processes over other surrogates is to provide an insight into the variance of the solution. The first method consists in using this data and weight it with the eigenvalues of the POD:
\begin{align}
\sum_{i=1}^k \sigma_i^2 \times \mathbb{V}[\mathcal{M}_{gp}(\mathbf{x}_*)]_{i}.
\end{align}

Global optimization of this indicator gives the new point to simulate~\cite{wales1997}.

\item Leave-One-Out (LOO) and $\sigma$, \hfill\\
A LOO is performed on the POD and highlights the point where the model is the most sensitive. The strategy here is to add a new point around it. The creation of the hypercube is described in \cref{sec:hypercube}. Within this hypercube, a global optimization over $\sigma$ is conduced giving the new point.

\item  LOO-\textit{Sobol'}, \hfill\\
Using the same steps as with the LOO-$\sigma$ method, the hypercube around the point is here truncated using prior information about \textit{Sobol'} indices---see \cref{sec:uq}. It requires that indices be close to convergence not to bias the result. However, the bias can be intentional depending on the insight we have about the case.

\item  Hybrid.\hfill\\
This last method consists of a navigator composed by any combination of the previous methods.
\end{itemize}

The evaluation of the latter composite method is not presented in this work. Although the computation of the LOO metric is merely an attempt to characterize the model's global quality, this mainly serves to assess the surrogate model's stability. If the model's response surface is not affected by the removal of a particular point, it is interpreted as a stability---or a non sensitivity---of the model to this action. This technique aims at stabilizing the model.

\subsection{Construction of the Hypercube}
\label{sec:hypercube}

\begin{figure}[h]
\centering
\includegraphics[width=0.8\linewidth,keepaspectratio]{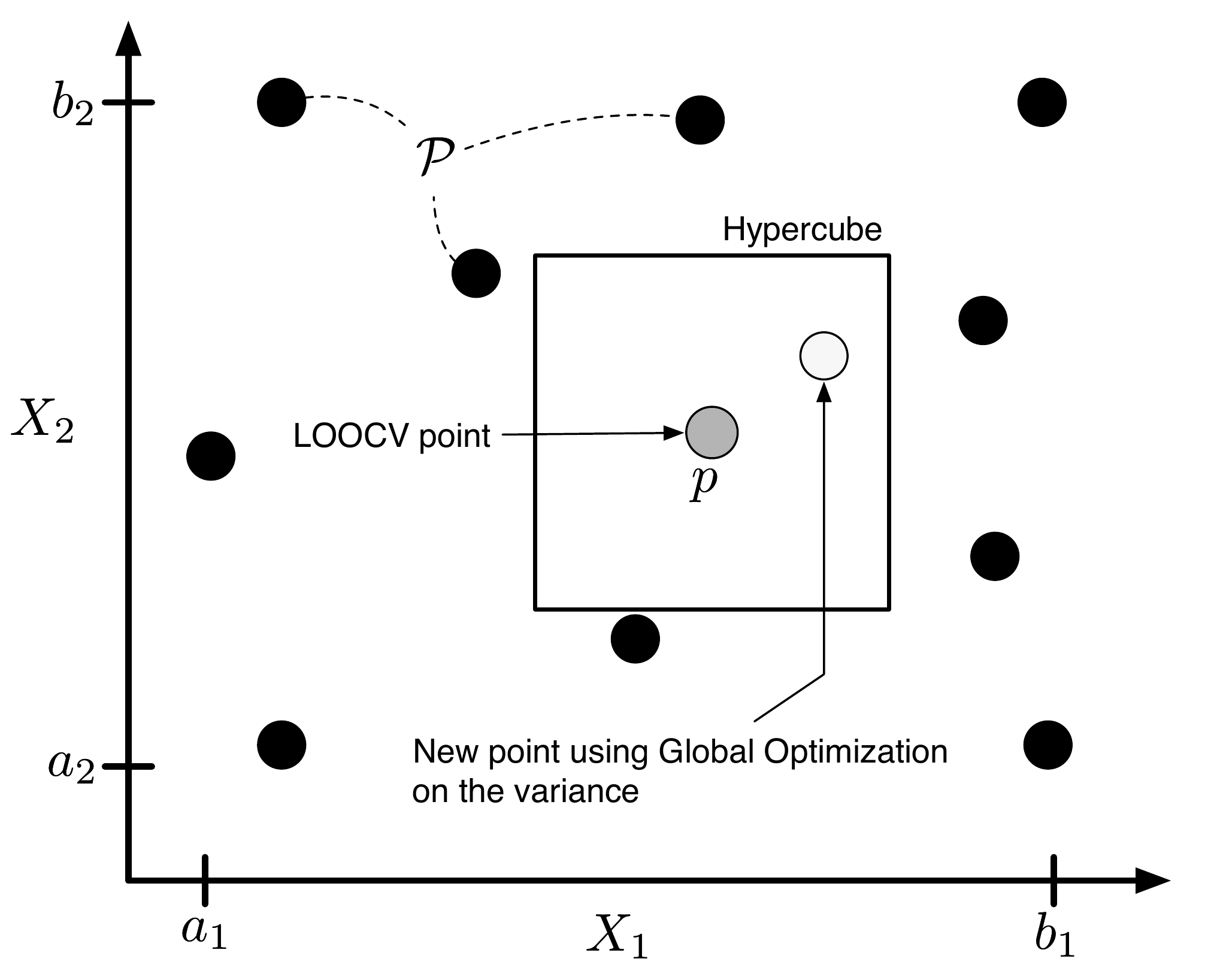}
\caption{Sketch of a Hypercube of size $[a_i, b_i]^2$. The grey dot is the LOO point $p$, the black dots are the surrounding points $\mathcal{P}$ and the white dot is the new point to evaluate.}
\label{fig:hypercube}
\end{figure}

To resample locally the parameter space, a hypercube is constructed around point $p$ which is the most sensitive in the construction of the surrogate model---LOO point, see \cref{sec:error}. An optimization problem is defined to construct the largest hypercube bounded by the surrounding points $\mathcal{P}$ as shown in \cref{fig:hypercube}. This allows to only consider the vicinity of the point.

The hypercube is defined by the cartesian product of the intervals of the $n$ parameters \textit{i.e.} $[a_i, b_i]^n$. The constrained optimization problem can hence be written as:
\begin{align}
\left\{\begin{array}{rc} \max  &\parallel (\mathbf{b} - \mathbf{a}) \parallel_{2} \\\mathcal{P} &\notin [a_i, b_i]^n \\ p &\in [a_i, b_i]^n \end{array}\right. .
\end{align}
A maximum cube-volume aspect ratio~\cite{smith1998} is also defined in order to preserve the locality. This gives the new constrain
\begin{align}
C : \sqrt[n]{\frac{\max (\mathbf{b} - \mathbf{a})}{\displaystyle\prod_{i = 1}^n \max (b_i - a_i)}} < \epsilon ,
\end{align}
with $\epsilon = 1.5$, set arbitrarily to prevent too elongated hypercubes. The global optimum is found using a two-step strategy: first, a discrete optimization using $\mathcal{P}$ gives an initial solution; second a basin-hopping algorithm~\cite{wales1997} finds the optimum coordinates of the hypercube. In case of the LOO-\textit{Sobol'} method, the hypercube is truncated using the total order \textit{Sobol'} indices.

\section{Results}
\label{sec:results}

The benefits and mechanisms of the methods are first evaluated on complex analytical functions. The chosen functions are defined in~\cref{sec:functions}. Then, the treatment of the parameter space's boundary is presented in~\cref{sec:delta-space}. Taking into account this issue, the analytical functions are tested in~\cref{sec:res-functions}. Finally, the methods are evaluated on a realistic application in~\cref{sec:ls89} with the LES of the LS89 test case~\cite{arts1990}.

\subsection{Analytical functions}
\label{sec:functions}

In order to test the new resampling methods, three analytical functions---see~\cref{tab:functions}---with increasing numbers of input dimensions are presented, namely: \textit{(i)} \textit{Rosenbrock} ; \textit{(ii)} \textit{Ishigami} ; and \textit{(iii)} \textit{g-function}~\cite{molga2005,ishigami1990,saltelli2007}. They all are widely used because they are nonlinear and nonmonotonic. Note that, similar results were obtained on other functions.

\begin{table*}[h]
\centering
\begin{tabular}{lll}
\toprule
Function&Hypercube & Definition \\
\cmidrule{1-3}
\textit{Rosenbrock} & $[-2.048, 2.048]^2$ & $
f(X_1, X_2) = \sum_{i = 1}^{d-1}[100(x_{i+1} - x_i^2)^2  +(x_i -1)^2].$\\
\textit{Ishigami} & $[-\pi, \pi]^3$ & $ f(X_1, X_2, X_3) = \sin X_1 + 7 \sin^2 X_2 + 0.1 X_3^4 \sin X_1. $\\
\textit{g-function} & $[0, 1]^4$ & $
f(X_1, X_2, X_3, X_4) = \prod_{i=1}^4 \frac{\lvert 4X_i - 2\rvert + a_i}{1 + a_i}, \quad a_{i} = i.$\\
\bottomrule
\end{tabular}
\caption{Analytical functions considered sorted by increasing number of input parameters.}
\label{tab:functions}
\end{table*}

\subsection{Restriction of the DoE}
\label{sec:delta-space}

\begin{figure}[h]
\centering
\includegraphics[width=\linewidth,keepaspectratio]{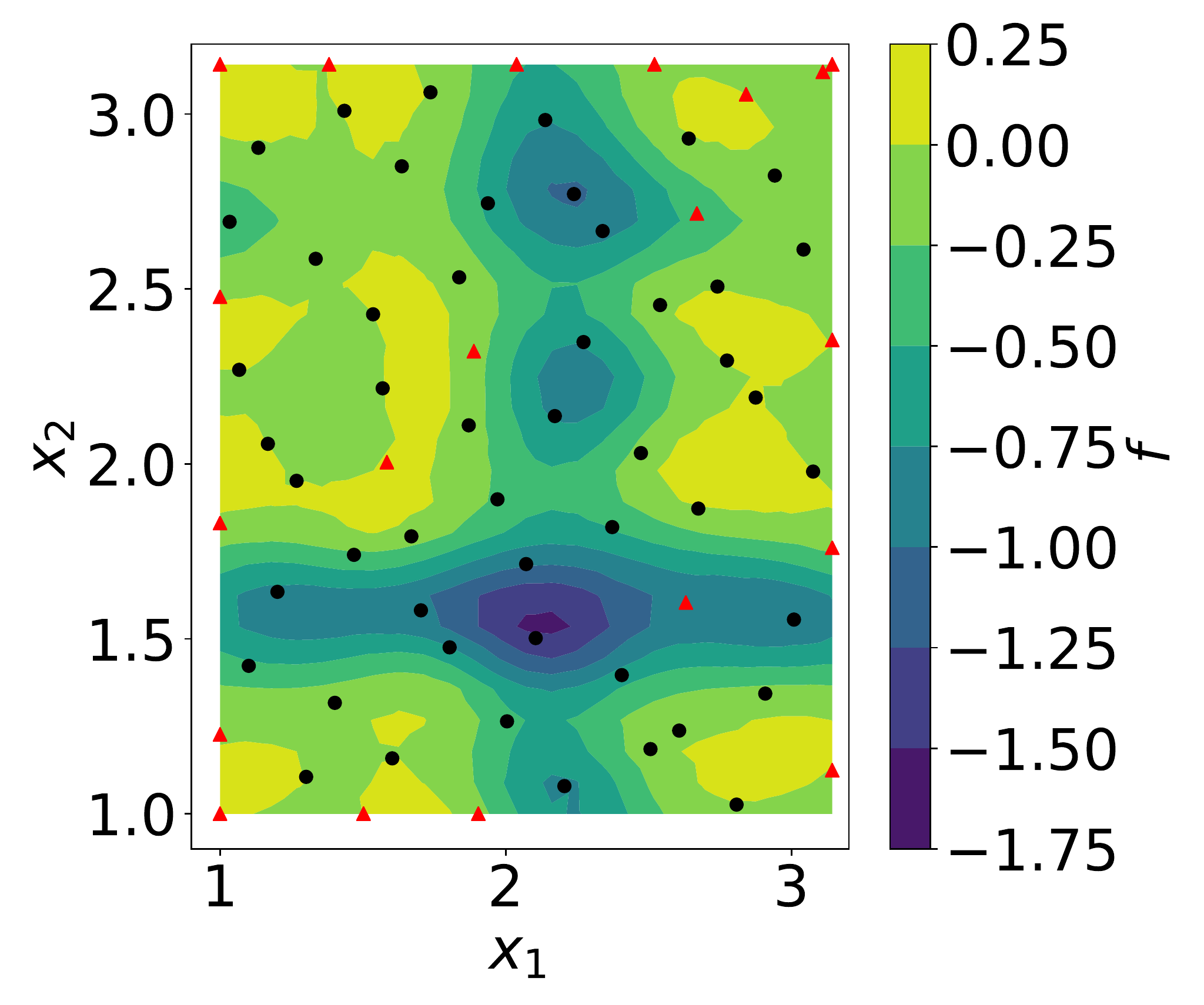}
\caption{\textit{Michalewicz} function: dots represent the initial sample of 50 points and diamonds represent the 20 resampled points. The function was evaluated on the hypercube $[1, \pi]^2$}
\label{fig:rs-michalewicz}
\end{figure}

The first step when constructing a model is to define the DoE. This is done by defining the range of each input parameter, the boundaries that describe a hypercube. Then, using a low discrepancy sequence as described in \cref{sec:doe}, an initial pool of snapshots is computed within the hypercube. However, when constructing a model based on Gaussian Process regression, the error is important at the boundaries of the DoE due to the lack of information. The model is thus not able to extrapolate accurately at these locations. If using the variance technique as it is, the algorithm tends to add points around the corners and only after it considers other parts of the domain. When dealing with a low dimensional case---fewer than three parameters as with the \textit{Michalewicz} function which uses two input parameters, see \cref{fig:rs-michalewicz}---, a few iterations are "wasted" in the process.

When increasing the number of parameters, there is a larger number of boundaries to cover. This has been confirmed on the \textit{Ishigami} function (3 input parameters) for which the reported $Q_2$ values are even worse. As shown in \cref{tab:size-q2-ishigami}, the optimization process is being over constrained in these regions and the global predictions are degraded. To obtain this Table, the initial sample was increased using a constant number of resampling points (10 points) and the error was measured using a uniform distribution on the domain, confirming the importance of the boundary treatment.

\begin{table}[h]
\centering
\begin{tabular}{lcc}
\toprule
Initial sample&Total size & $Q_2$ \\
\cmidrule{1-3}
30 & 40 & 0.05  \\
35 & 45 & -0.02 \\
40 & 50 & -0.13 \\
45 & 55 & -0.19 \\
50 & 60 & -0.04 \\
55 & 65 & 0.43  \\
60 & 70 & 0.51  \\
65 & 75 & 0.87  \\
70 & 80 & 0.54  \\
75 & 85 & 0.86  \\
\bottomrule
\end{tabular}
\caption{Error $Q_2$ on the \textit{Ishigami} function of the size of the initial sample using a variance strategy with 10 points.}
\label{tab:size-q2-ishigami}
\end{table}

\begin{figure}[!h]
\centering
\includegraphics[width=0.8\linewidth,keepaspectratio]{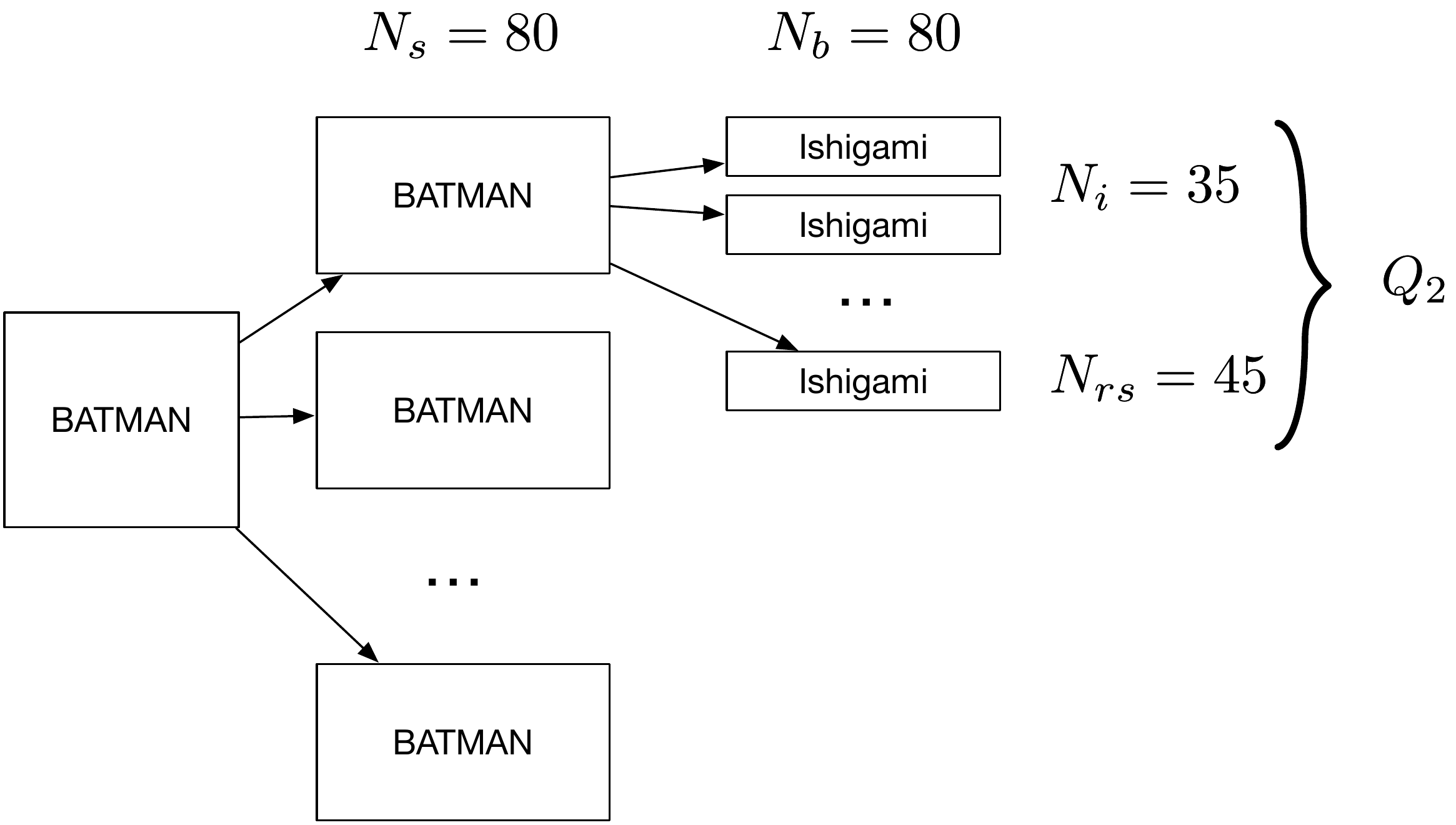}
\caption{Example showing a computation of $Q_2$ with $N_i = 35, N_{rs} = 45$.}
\label{fig:jpod-delta}
\end{figure}

The possibility to widen the space by a delta space has been evaluated to address this question. The objective is to condition the predictor around the boundaries by adding information outside the domain of interest. A Halton sequence has been used to generate a sample of size $N_s =80$ from the space
\begin{align}
N_i\sim\mathcal{U} (20, 80) \quad \Delta_{space}\sim\mathcal{U} (0, 20\%),
\end{align}
with $N_i$ the number of initial snapshots and $\Delta_{space}$ the widening factor, the outer delta space. For each case $N_{i}$, it is only the proportion of initial sample over the number of resample point that varies. (See~\cref{fig:jpod-delta}). A fixed budget of $N_b = 80$ snapshots was considered. Then, the number of resampling points is equal to $N_{rs} = N_b - N_i$. The strategy used here was the $\sigma$ model (see~\cref{sec:doe}). After the resampling phase has been completed, the quality $Q_2$ of the model is computed. Applied to the \textit{Ishigami} function, $N_s$ simulations each performing $N_{b}$ evaluations have been used to construct the response surface. These results were compared to a case without resampling: $N_i = N_S = 80$. The resulting predictivity quality being $Q_2\simeq0.8$.

\begin{figure*}[h]               
\centering
\subfloat[Uniform distribution]{
\includegraphics[width=0.47\linewidth,height=\textheight,keepaspectratio]{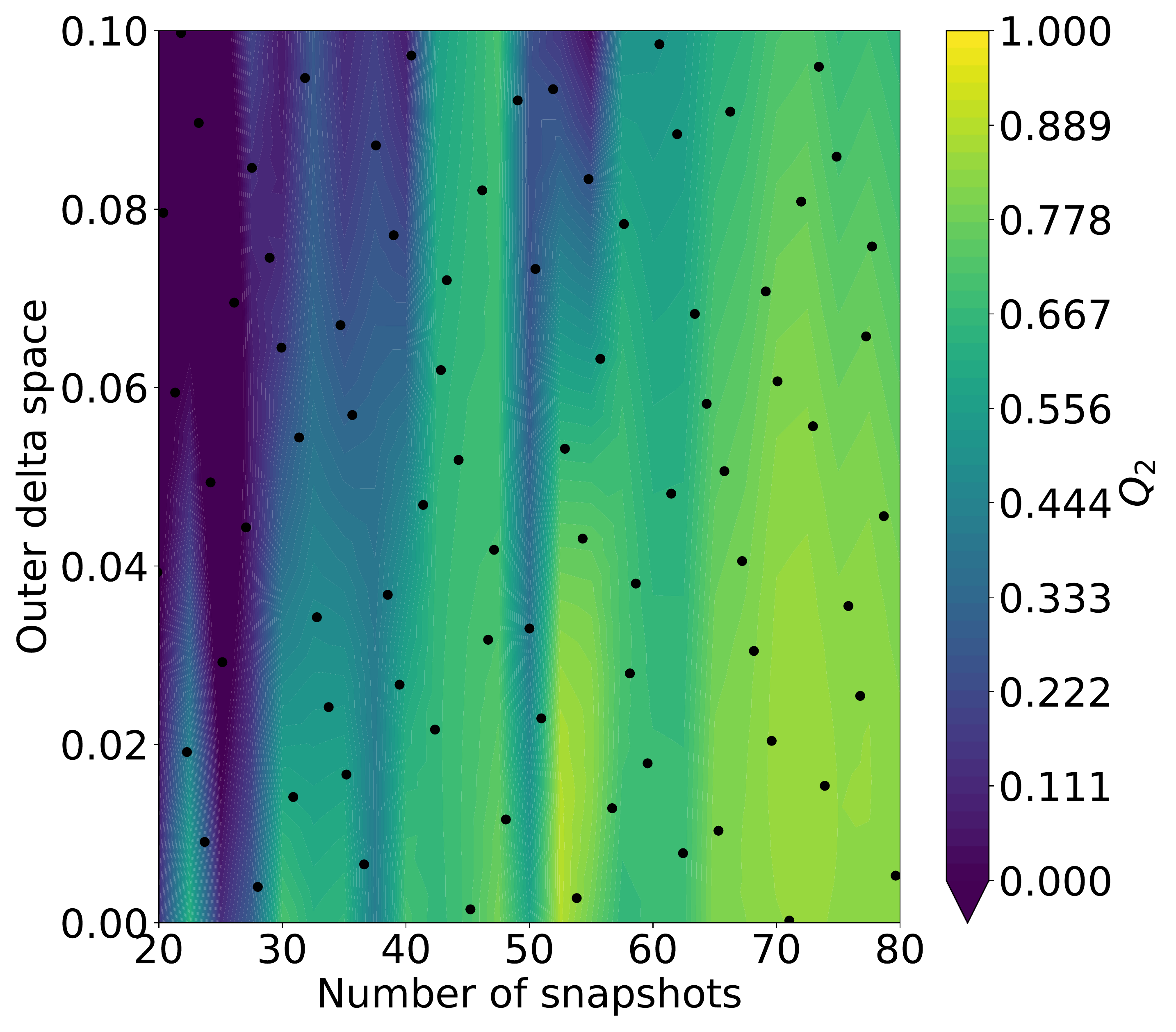}}
 ~       
\subfloat[Arcsine distribution]{
\includegraphics[width=0.47\linewidth,height=\textheight,keepaspectratio]{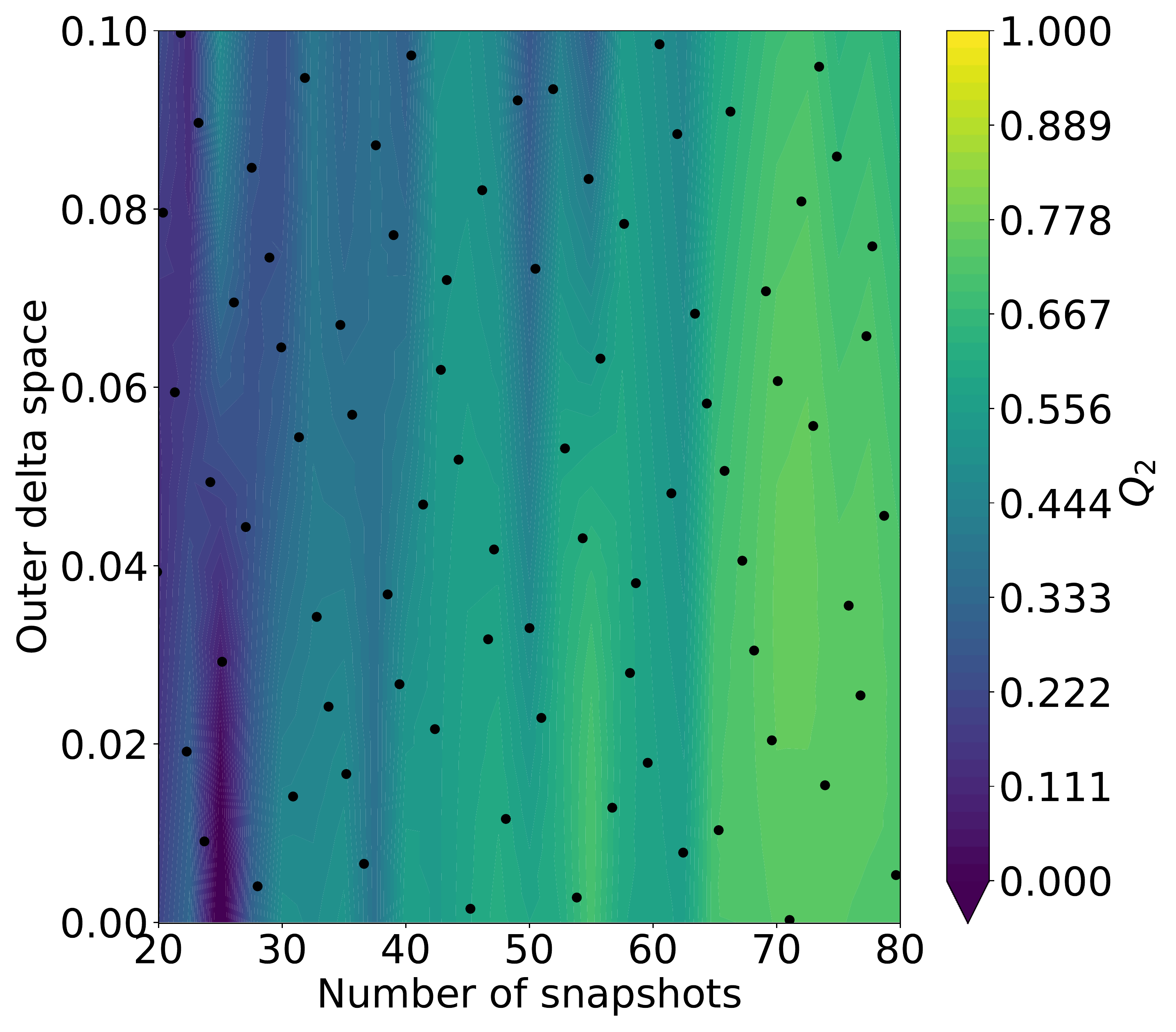}}
\caption{Response surface of $Q_2$ function of the initial sample and the \textit{outer delta space}. Dots represent the simulations.}
\label{fig:outer-delta}
\end{figure*}

As shown in~\cref{fig:outer-delta}, there is no benefit of adding points outside the domain. Aside from the uniform distributions usually employed on this function, a standard arcsine distribution was also tested to assess the quality around boundaries but no enhancement was observed. When the delta space is increased, there is a loss of quality due to the presence of points in non-interesting regions.

\begin{figure*}[h]               
\centering
\subfloat[Uniform distribution]{
\includegraphics[width=0.47\linewidth,height=\textheight,keepaspectratio]{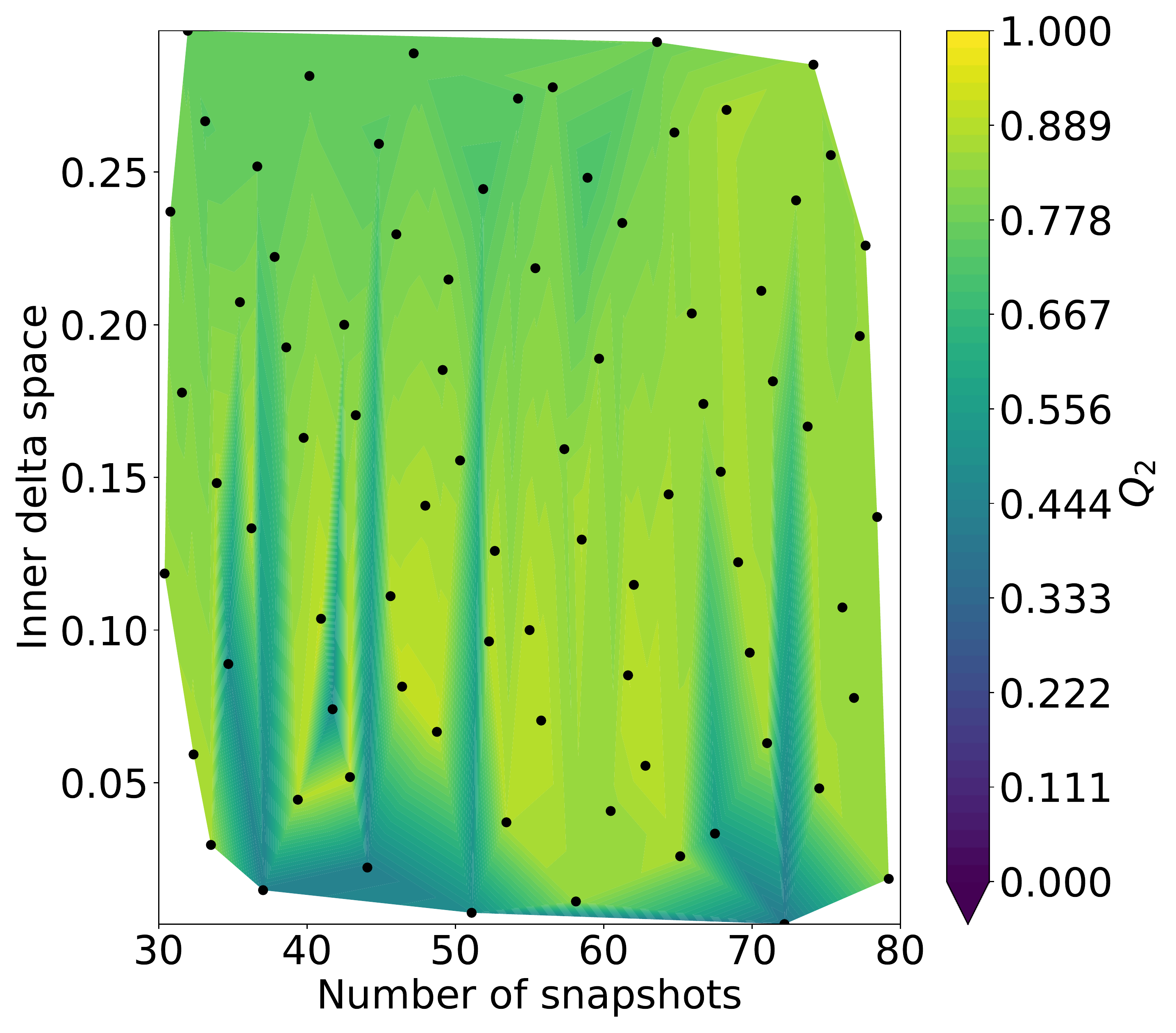}}
 ~       
\subfloat[Arcsine distribution]{
\includegraphics[width=0.47\linewidth,height=\textheight,keepaspectratio]{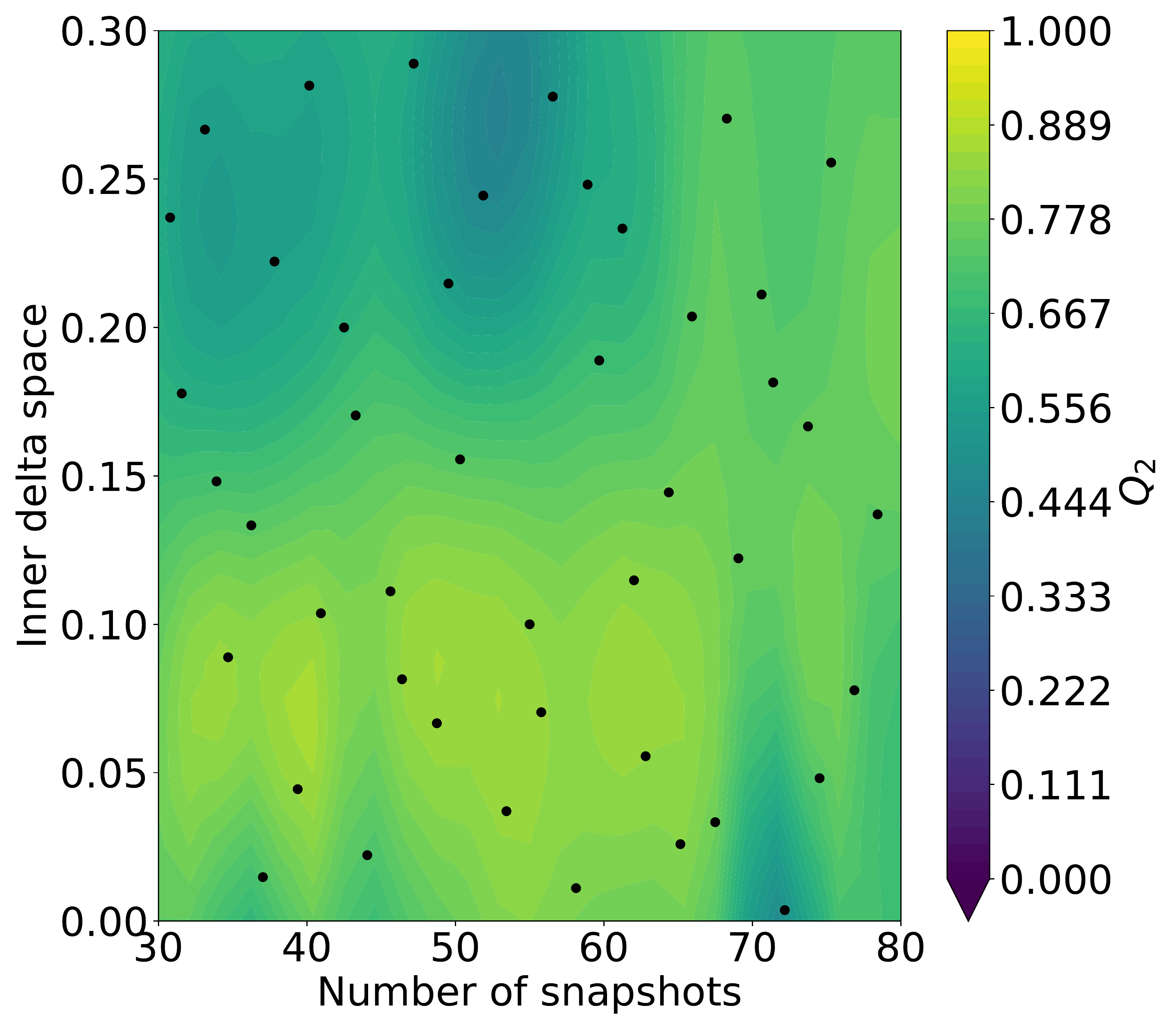}}
\caption{Response surface of $Q_2$ function of the initial sample and the \textit{inner delta space}. Dots represent the simulations.}
\label{fig:inner-delta}
\end{figure*}

Complementarily to this analysis using an outer delta space, an inner delta space factor has also been considered. The same methodology was used. Results are shown in \cref{fig:inner-delta}. On the uniform case, the model was not correctly computed due to high discontinuities caused by the ~0\% inner delta space cases. In~\cite{dette2010}, optimal design that tends to put more points near the boundaries were shown to be more effective. Our results are coherent with their findings as we observed an improvement of the quality when using a low inner delta space. Indeed, a small value of the parameter limits the trend to add points close to the boundaries.

This work has shown that setting an inner delta space comprised between 5 and 10\% is required to ensure the robustness of the model construction. Based on this observation, in the following the inner delta space is set to an arbitrary value of 8\%.

\subsection{Application on analytical functions}
\label{sec:res-functions}

\begin{figure*}[!h]               
\centering
\subfloat[$\sigma$: $Q_2 = 0.75$]{
\includegraphics[width=0.47\linewidth,height=\textheight,keepaspectratio]{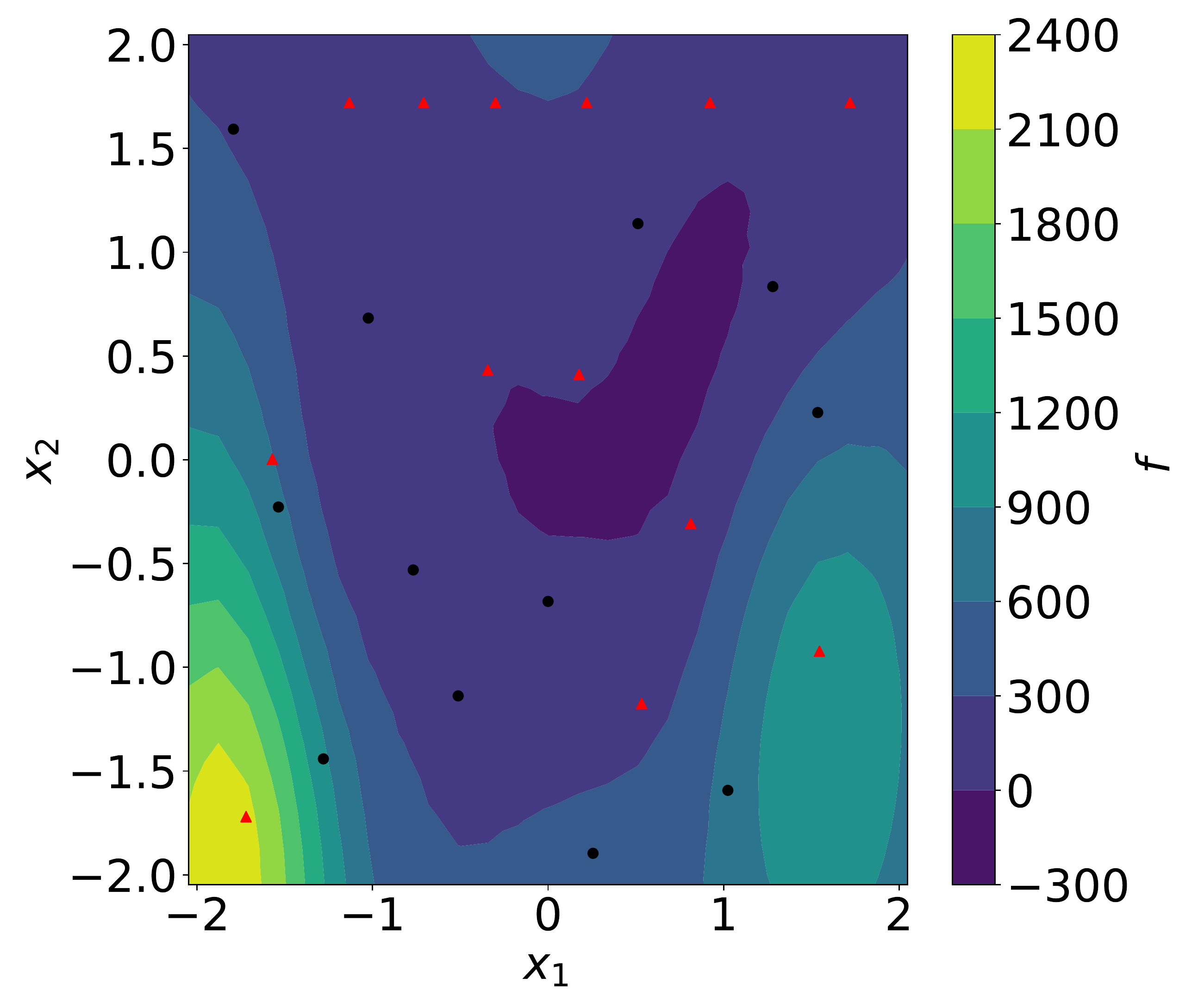}}
 ~       
\subfloat[LOO-$\sigma$: $Q_2 = 0.68$]{
\includegraphics[width=0.47\linewidth,height=\textheight,keepaspectratio]{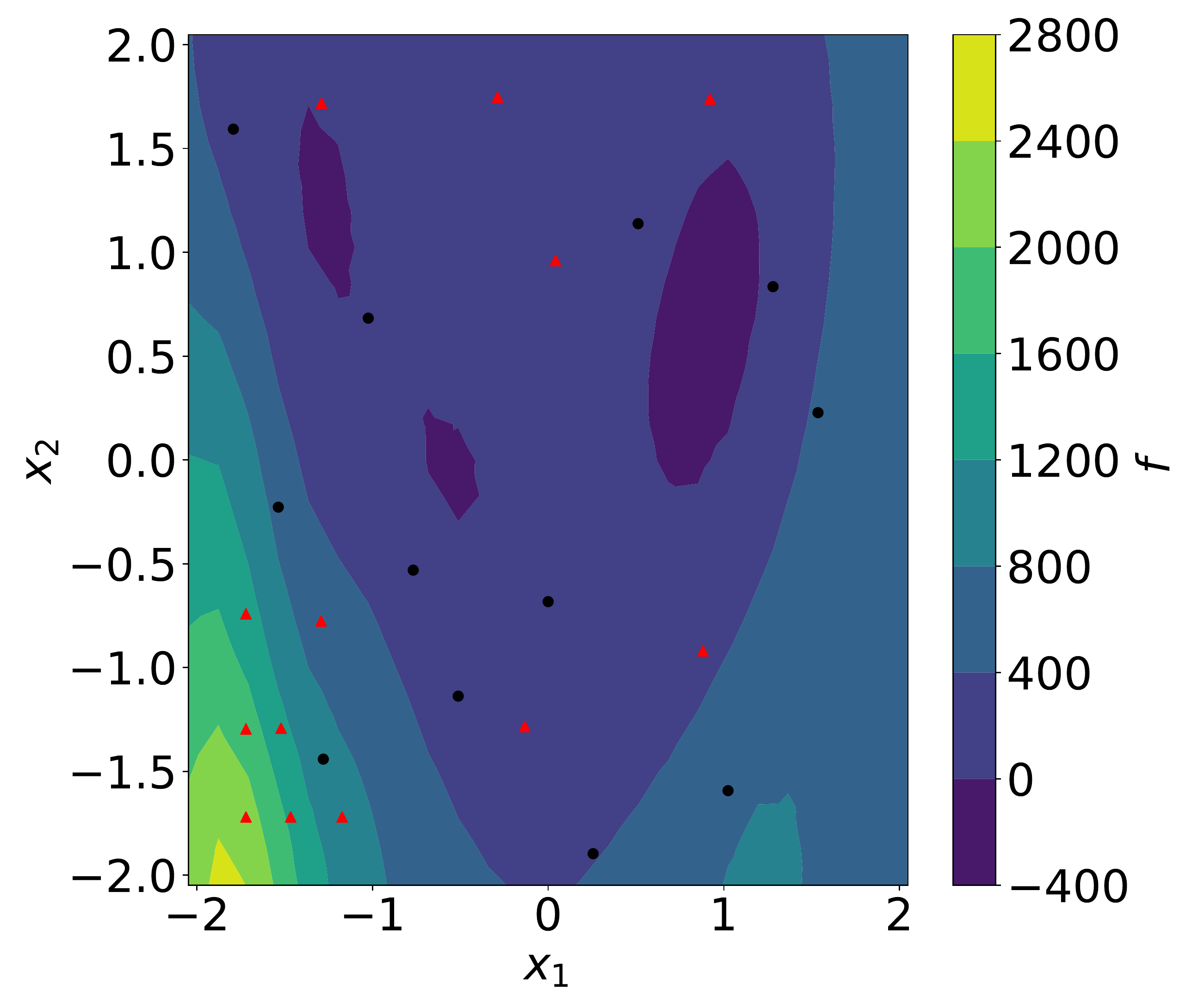}}
      
\subfloat[LOO-\textit{Sobol'}: $Q_2 = 0.86$]{
\includegraphics[width=0.47\linewidth,height=\textheight,keepaspectratio]{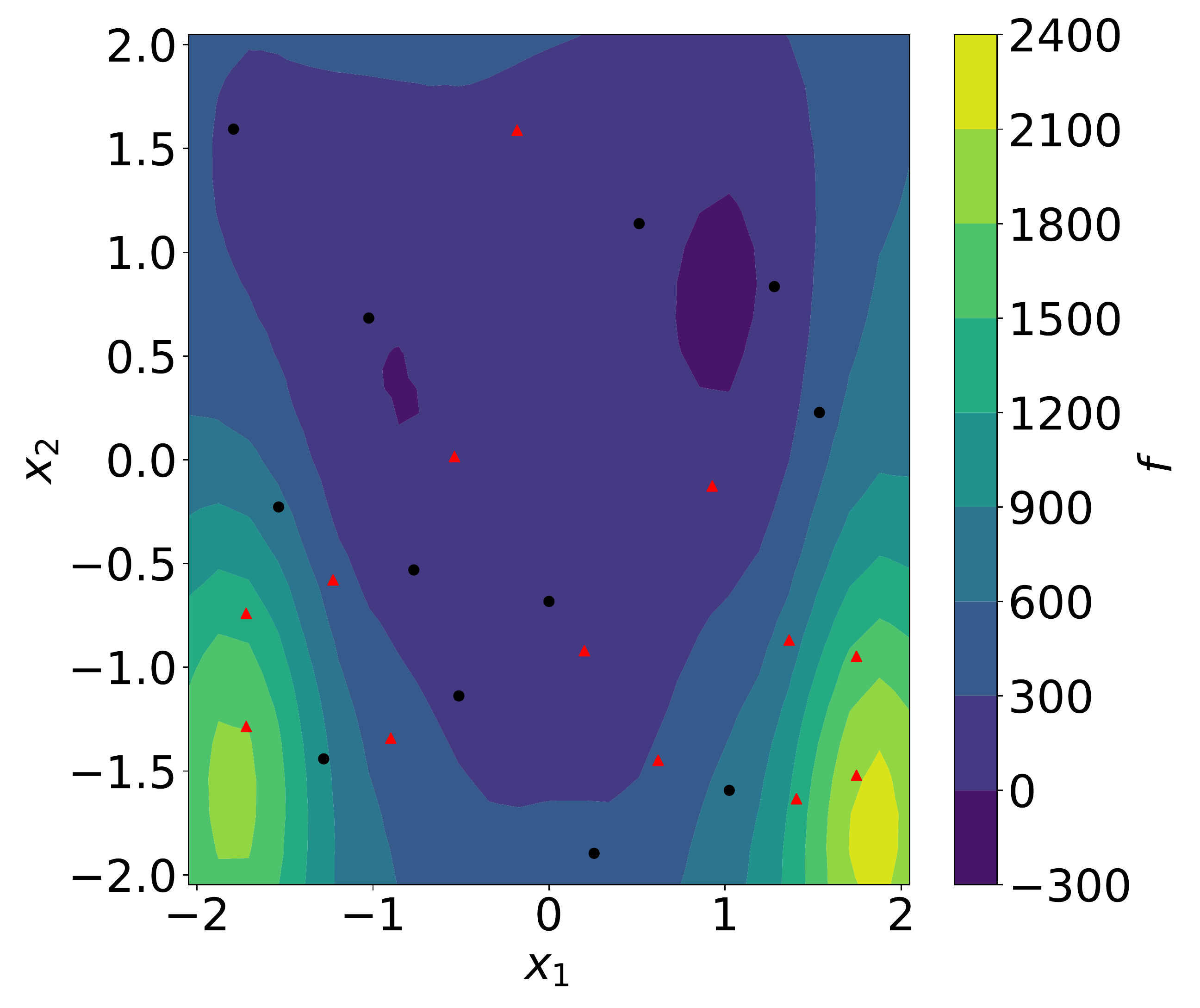}}
\caption{Response surface of the \textit{Rosenbrock} function. In each case, the initial learning sample is composed of 12 simulations and there are 13 resampling points---respectively represented in dots and diamonds.}
\label{fig:methods-rosenbrock}
\end{figure*}

The operating mechanism and catches of the method can be visualized on the \textit{Rosenbrock} function---see~\cref{fig:methods-rosenbrock}. Starting from the $\sigma$ method: points are first added close to the top boundary despite the inner delta space parameter. However, the lack of surrounding points made this choice fairly legitimate. Other points seem to be located in interesting regions---where there is a gradient and no points. It can be seen as a low discrepancy sequence, which made its use relevant for studying the delta space impact in \cref{sec:delta-space}. On the other hand, the LOO-$\sigma$ method does not seem to exhibit a boundary preference. But, on the bottom left-hand corner, there is an accumulation of points. Indeed, this method relies on the location of the most sensitive point. Considering the surroundings of a strong extremum---as it is the case here---, the method tends to add points first in this zone preventing further exploration of the domain and, in this case, totally misses the second extremum. Lastly, the LOO-\textit{Sobol'} method seems more balanced. Points have been added preferentially on the $X_1$ parameter axis, as it is slightly the most influent parameter ($S_{T_{X_1}} \simeq 0.7$). 

A convergence study has also been performed. With a fixed total number of simulations, the size of the initial learning sample was changed to evaluate the impact of the ratio of the initial sampling over the total number of samples on the quality of the model. As in~\cref{sec:delta-space}, a \textit{Halton} sequence was used. The respective parameters are reported in~\cref{tab:cv-budget-q2}. The \textit{Sobol'} indices for the \textit{Ishigami} function are found in~\cite{marrel2012}, while for the other functions, a deterministic sample of $\numprint{100000}$ evaluations was used.

\begin{table*}
\centering
\begin{tabular}{lccl}
\toprule
Function &Sample Budget & $Q_2$ & Total order \textit{Sobol'} indices \\
\cmidrule{1-3}
\textit{Rosenbrock} 2-D & 25 & 0.82 & [0.71, 0.50]\\
\textit{Ishigami} 3-D& 80 & 0.85 & [0.557, 0.443, 0.244]\\
\textit{g-function} 4-D& 65 & 0.66& [ 0.61,  0.29,  0.16,  0.09]\\
\bottomrule
\end{tabular}
\caption{Reference $Q_2$ and Total order \textit{Sobol'} indices at a fixed budget using a Halton sequence for the analytical functions.}
\label{tab:cv-budget-q2}
\end{table*}

\begin{figure*}[!h]
\centering
\subfloat[\textit{Rosenbrock}]{
\includegraphics[width=0.47\linewidth,height=\textheight,keepaspectratio]{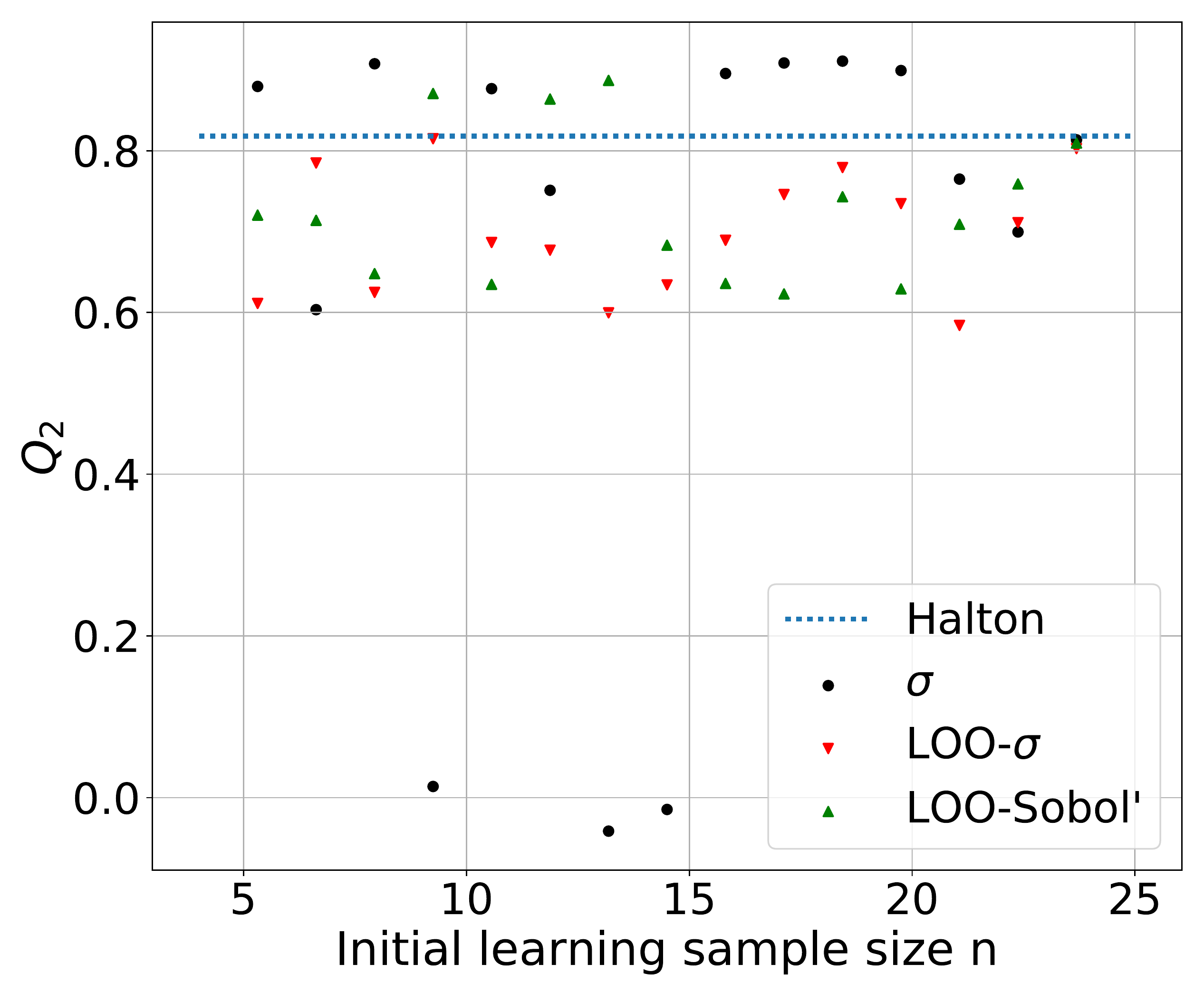}}
 ~       
\subfloat[\textit{Ishigami}]{
\includegraphics[width=0.47\linewidth,height=\textheight,keepaspectratio]{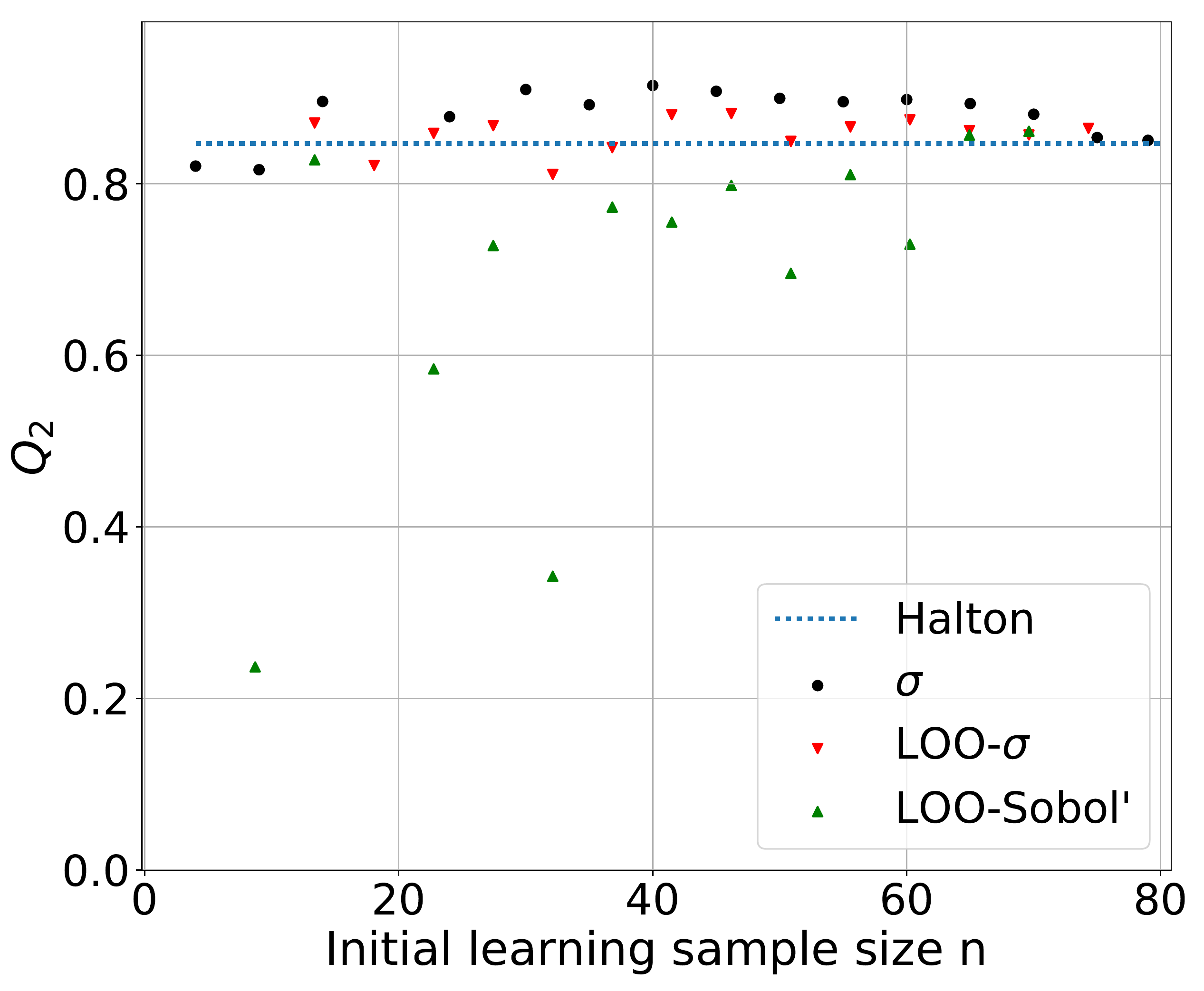}}
        
\subfloat[\textit{g-function}]{
\includegraphics[width=0.47\linewidth,height=\textheight,keepaspectratio]{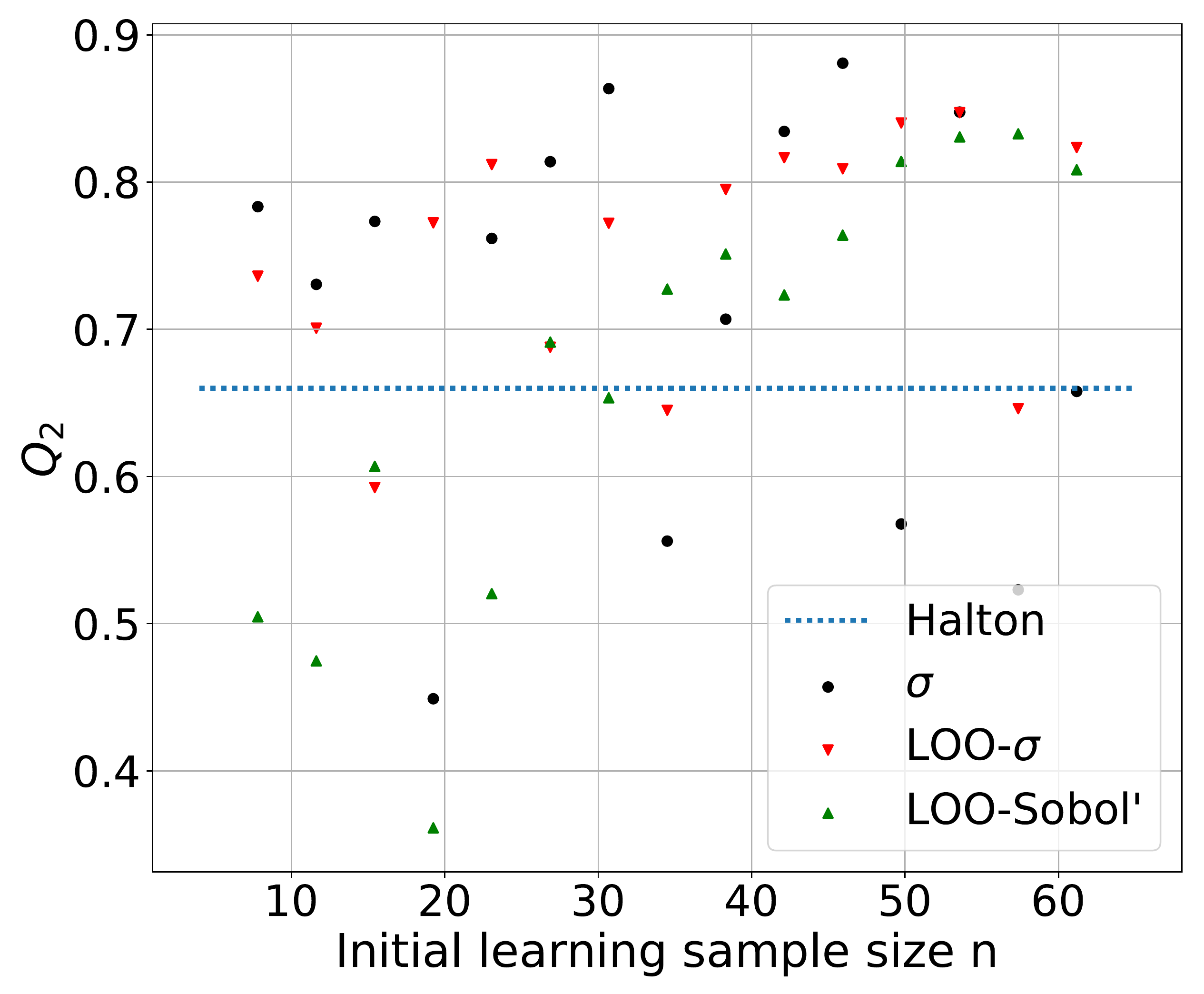}}
\caption{Convergence of $Q_2$ of the different methods on each function by varying the initial learning sample size with a fixed budget.}
\label{fig:methods-cv}
\end{figure*}

\begin{figure*}[!h]               
\centering
\subfloat[Gaussian Process: $Q_2 = 0$]{
\includegraphics[width=0.47\linewidth,height=\textheight,keepaspectratio]{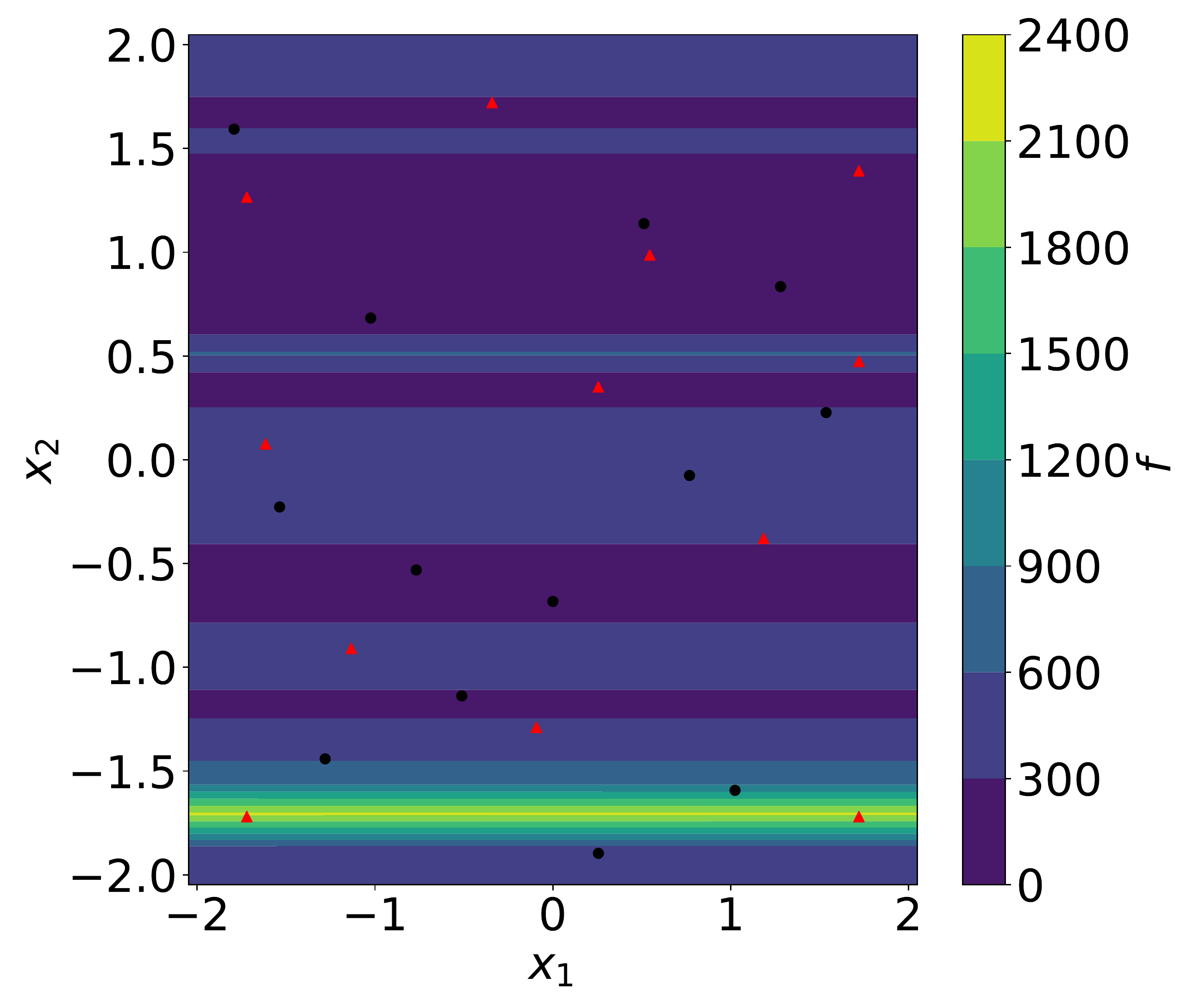}}
 ~       
\subfloat[RBF: $Q_2 = 0.83$]{
\includegraphics[width=0.47\linewidth,height=\textheight,keepaspectratio]{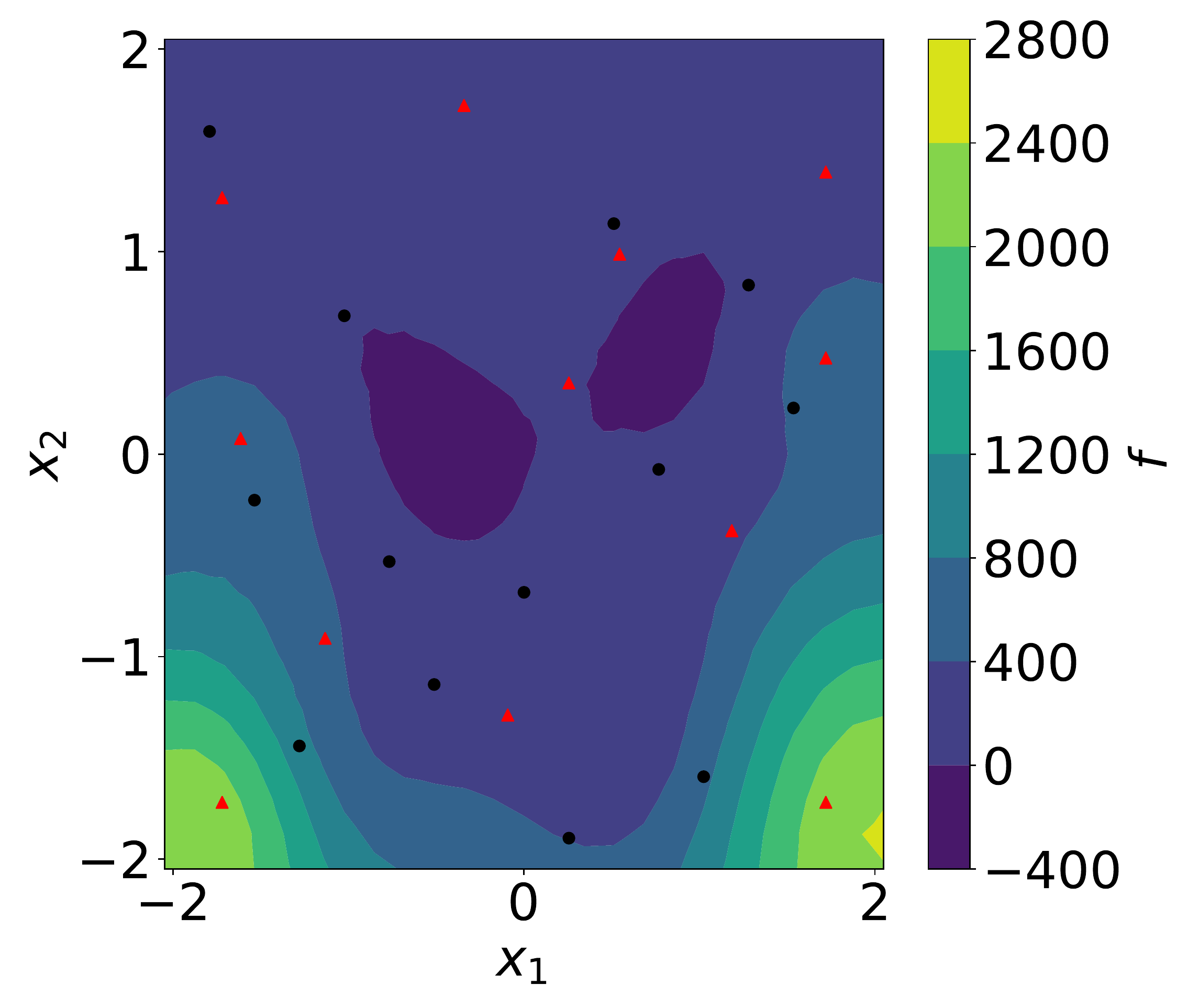}}
\caption{Response surface of the \textit{Rosenbrock} function. Comparison between two models. The initial sample is composed of 13 simulations and 12 resampling points---respectively represented in dots and diamonds.}
\label{fig:methods-rosenbrock-nofit}
\end{figure*}

Results are shown in~\cref{fig:methods-cv}. The $\sigma$ method appears to be one of the most, in some cases the most, effective method but it also exhibits more variability. Increasing dimensionality seems only to improve slightly this behaviour. There are multiple explanations to this phenomenon. The method relies on the use of an inference about the variance of the model. Starting from a given sample, if the fitting process does not converge, the prediction of the variance will be far from correct leading to a wrong resampling. Of course, there is a chance for this new point location to be relevant, still this can lead to an even worse model or an overfitting where the model is too closely linked to the outputs, so the model has memorized only the feature but not learned the underlining correlation between the data. Lastly, looking at~\cref{fig:methods-rosenbrock-nofit}, even if the points look well distributed over the parameter space, the pGP model is absolutely wrong. The Gaussian Process reconstruction failed to recover the response surface of the function whereas a Radial Basis Function Networks model successfully did it.

The other two methods share the $\sigma$ strategy, but the variability is conditioned by the LOO point. Indeed, the former only uses inference about the predictive variance whereas LOO's methods take into account the observed quality of the model.
LOO-\textit{Sobol'} is even more stable especially when the contribution of the parameters to the QoI is not even. The quality evolves quasi-linearly with the initial sample size. This is due to the initial guess on the indices. The closer the indices are converged, the better the sizing of the hypercube used by the $\sigma$ strategy. Indeed, some dimension of the hypercube could be neglected due to the indices. In the \textit{Rosenbrock} case the method behaves like LOO-$\sigma$, the importance factors are close enough so that this collapse of dimension does not occur. On the other hand, with the \textit{g-function}, the total order \textit{Sobol'} indice of the last input parameter is so small that the algorithm tends not to take into account this dimension.

For each function, as the initial sample gets close to the budget, the expected improvement is reduced. This is clear with the \textit{Ishigami} function. When the initial sample is too small, the model is so poor that the points are not added efficiently. On the contrary, if we add an insufficient number of points, the impact is close to none but still there is an improvement. From the other cases, the effect of the ratio of the initial learning sample size over the total budget is not so clear. In 2-D the impact is null and after that, a ratio $>0.5$ seems appropriate.

Thus, setting aside the possible non-fitting of the data, improving the quality of the surrogate model by resampling the parameter space appears to be guaranteed in high dimensional cases and using no more than half of the budget.

\subsection{LS89 case}
\label{sec:ls89}

\subsubsection{Case description} 
The \textit{LS89} case is a blade cascade designed and tested experimentally at the Von Karman Institute for Fluid Dynamics (VKI)~\cite{arts1990}. The linear cascade consists of five high-pressure turbine vanes although only the center vane is studied. The vane is a 2D extruded profile unlike most industrial vanes that are much more complex geometrically. It however remains of great interest because the operating points are representative of values found in real engines today. This test case represents one of the largest turbomachinery databases available for the validation of CFD models in complex geometries.

A large variety of operating points have been successfully simulated until now. Low levels of turbulence injection ($<1$\%) do not represent an issue for most solvers \cite{Gourdain2010,emory2016} using either Reynolds-Averaged Navier Stokes (RANS) or Large Eddy Simulation (LES). Higher levels of turbulence have also been studied successfully~\cite{Wheeler2015} but difficulties arise for higher Reynolds numbers and larger outlet Mach numbers. Simulations are not able to correctly predict experimentally obtained profiles, notably the heat transfer field which is of great importance for the blade life-cycle.

\begin{figure*}[!h]
\centering
\includegraphics[width=0.8\linewidth,keepaspectratio]{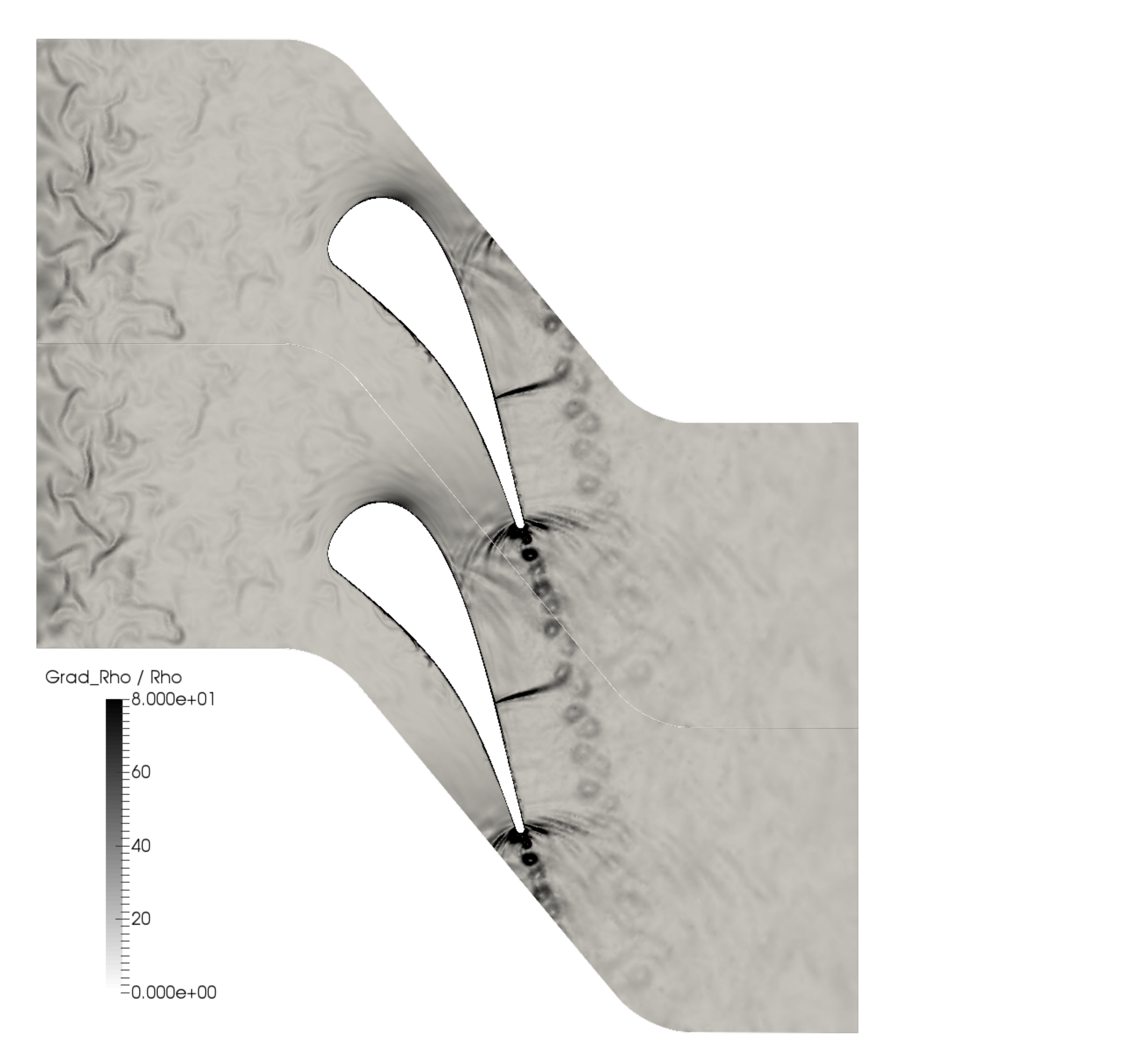}
\caption{$\frac{\nabla \rho}{\rho}\; (m^{-1})$ with $Tu = 30\%$.}
\label{fig:ls89-gradRhoRho}
\end{figure*}

The operating point addressed in this document, selected from Arts~\cite{arts1990}, is the MUR235, a very rich case in terms of physics that presents the above mentioned challenges (high Reynolds and outlet Mach numbers). Figure~\ref{fig:ls89-gradRhoRho} highlights the main physical interactions in such a flow. One of the most notable features is the presence of a shock wave on the suction side of the blade. This shock wave interacts with a transitional boundary layer due to the highly curved flow, a potential source of instabilities in the boundary layer which in turn determines the wake downstream. This wake issues acoustic waves that impact the neighbour blade affecting the stability of the boundary layer. Also, there is a high level of free-stream turbulence that undergoes stretching around the leading edge of the blade which modifies the position of the boundary layer transition on the suction side~\cite{Segui2017a}.


In the original experiments~\cite{arts1990}, an increase in heat transfer is observed on the suction side of the blade when a high turbulence intensity level at the inlet ($\sim 6$\%) as well as a large Reynolds number at the outlet ($>\numprint{1e6}$) are present. The simulations recover the shock wave that triggers an abrupt transition of the boundary layer, but turbulent spots may be found upstream of this position that can contribute to the overall heat transfer. These spots can be explained due to perturbations in the free-stream turbulence $Tu$ that are capable of trespassing the sheltering effect of the shear layer and thereby increase the heat transfer. Turbulence values upstream of the blade are thus of upmost importance.

The original experiments give only the turbulence intensity level at an upstream distance from the vane, which is insufficient to characterize the turbulent flow at this location. Recent studies on the same test bench have measured the integral length scale for the same intensity level~\cite{Fontaneto2014}. In spite of this newly available information, simulations are not capable of recovering an important part of the heat flux on the suction side even when taking the correct length scale~\cite{Pichler2016}. Uncertainties concerning the measured values in the experiments, that serve as boundary conditions in the simulation, appear as a path to be explored. 

\begin{figure*}[!h]
\centering
\includegraphics[width=0.9\linewidth,keepaspectratio]{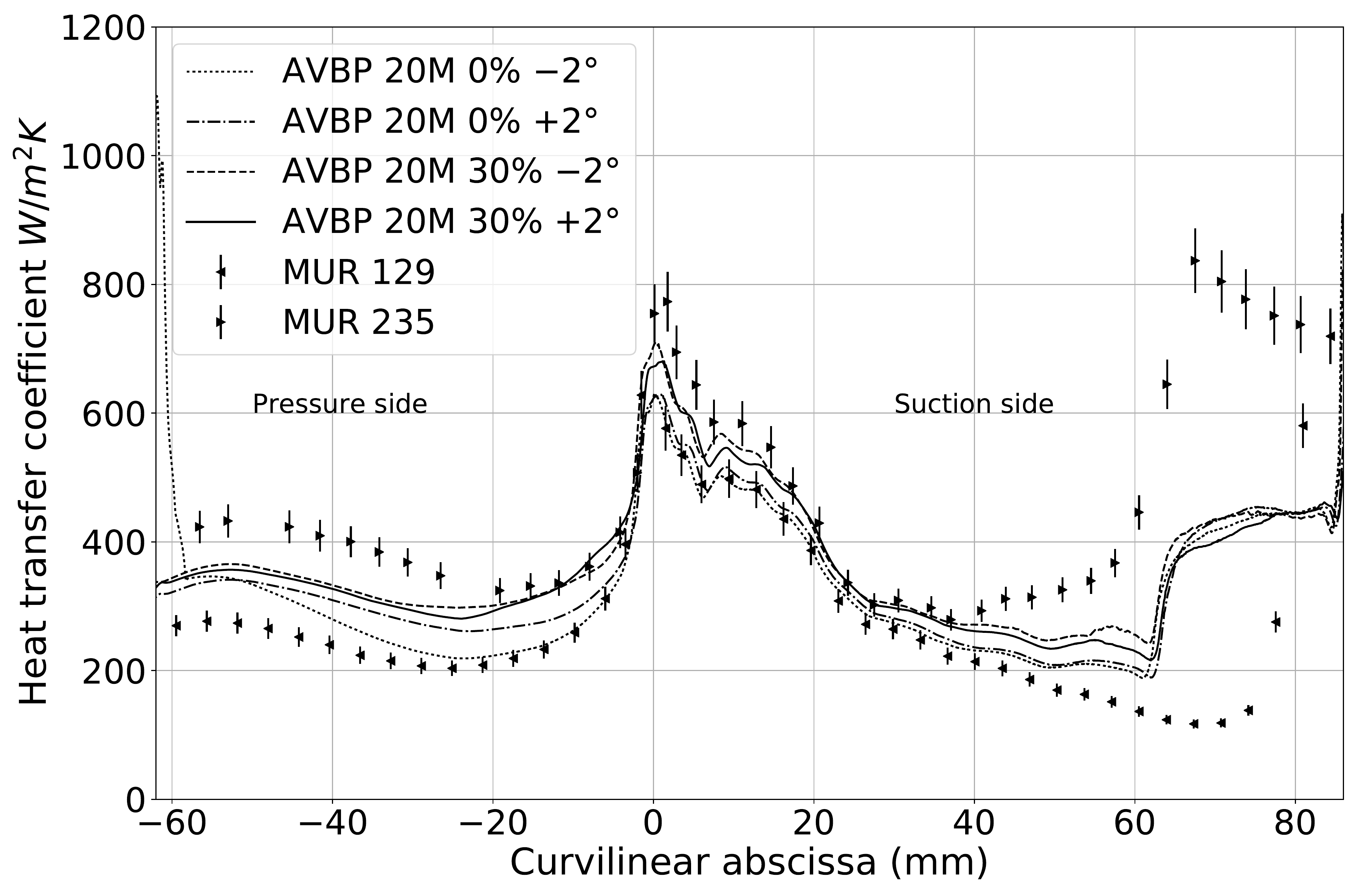}
\caption{Heat transfer coefficient variation compared to experimental data of MUR129 ($Tu=1\%, \alpha = 0\degree$) and MUR235 ($Tu=6\%, \alpha = 0\degree$).}
\label{fig:space-tu-alpha}
\end{figure*}

Apart from the turbulence intensity and the length scale, the angle of attack $\alpha$ of the incoming flow can also be seen as an uncertain parameter. There is no information related to this parameter in the experimental campaigns. In~\cref{fig:space-tu-alpha}, the effect of $\alpha$ was numerically investigated with respect to $Tu$ by studying the heat transfer coefficient response---hereafter defined as the QoI. Due to the computational effort required to modify and simulate correctly a case with a modified integral length scale versus a modification of $\alpha$, this parameter was not taken into account. Increasing $Tu$ or $\alpha$ causes an increase of the QoI and $Tu$ seems to have a larger impact than $\alpha$. A deeper analysis would require more computations to obtain: \textit{(i)} a correct response of the influence of these parameters on the QoI ; \textit{(ii)} the contribution of each parameter ; and \textit{(iii)} the probability density function of the QoI by propagating the uncertainties. Thus, the parameter space for this study was defined as
\begin{align}
Tu \in [0, 30\%] \quad \alpha \in [-5,5\degree].
\end{align}

\subsubsection{Numerical setup}

The simulations have been performed using \textit{AVBP}~\cite{Gicquel2011}, a validated CFD LES solver co-developed by CERFACS and IFP-EN. This parallel code solves the three-dimensional compressible Navier-Stokes equations for both steady and unsteady reacting flows. The code is capable of handling hybrid unstructured meshes and allows to address complex geometries. High-order numerical schemes based on the \textit{Taylor-Galerkin} (TTG) family are used~\cite{Quartapelle1993}. 

The simulations were performed on a 20 million cells mesh. Five layers of prisms in the near-wall region are present allowing a higher aspect ratio. The mean $\overline{y^+}$ has a value of $\simeq 6.62$ which limits the physical time step to $\unit{\numprint{1.94e-8}}{\second}$. In this context, a wall-resolved computation using the \textit{WALE} \cite{Nicoud1999} model is used to take into account the proper turbulence scaling in the near-wall region. To gather enough statistics, a simulation time of $\unit{\sim\numprint{4.1}}{\milli\second}$ was performed. This lead to a CPU cost, for a single computation, of $\sim7500$~hours lasting $\sim5$~hours on a cluster of 1440 cores.

\begin{figure*}[!h]
\centering
\includegraphics[width=0.9\linewidth,keepaspectratio]{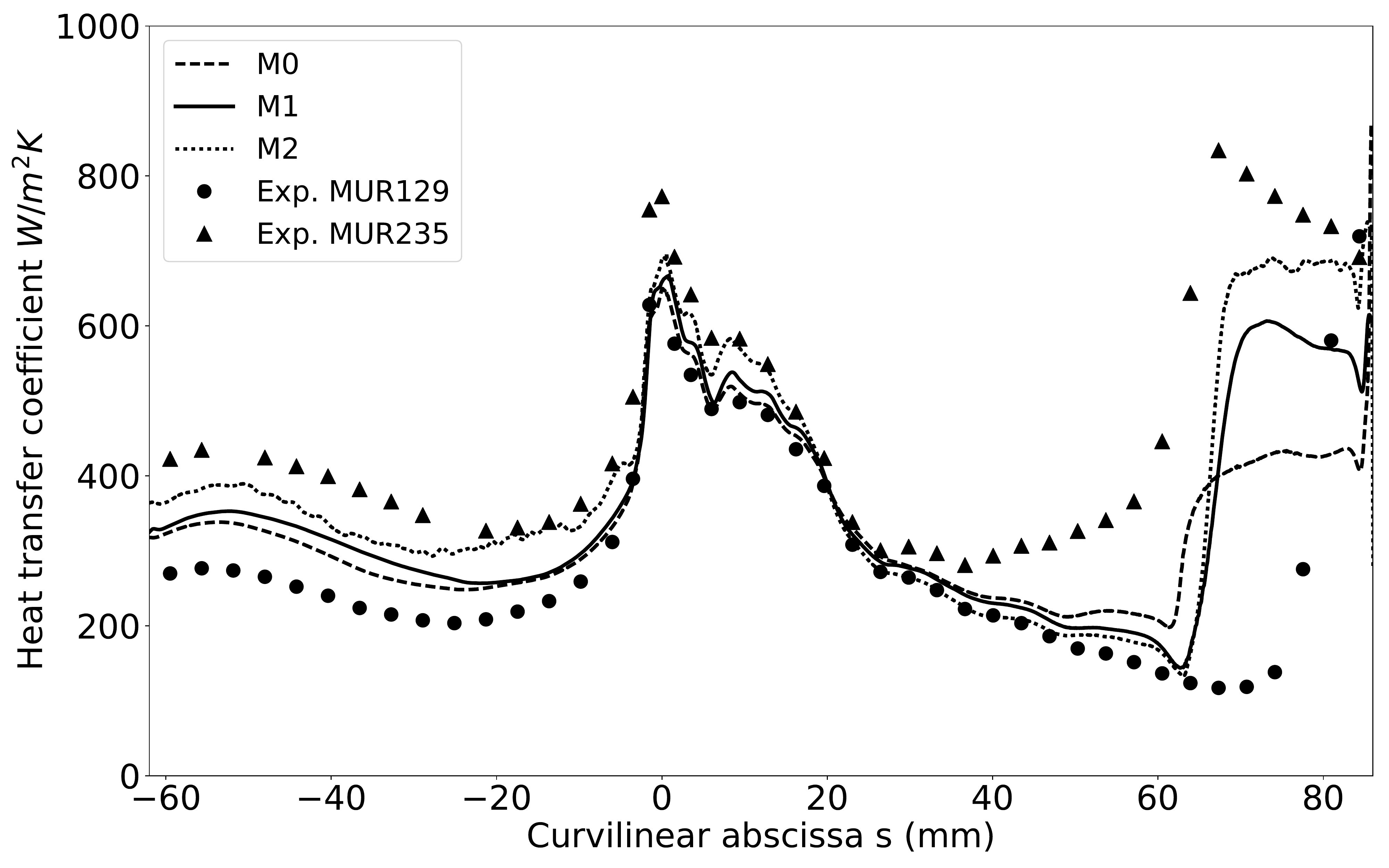}
\caption{Heat transfer coefficient between various meshes using MUR235 setup ($Tu=6\%, \alpha = 0\degree$).}
\label{fig:M0M1M2_hcomp}
\end{figure*}

\begin{figure*}[!h]
\centering
\includegraphics[width=0.9\linewidth,keepaspectratio]{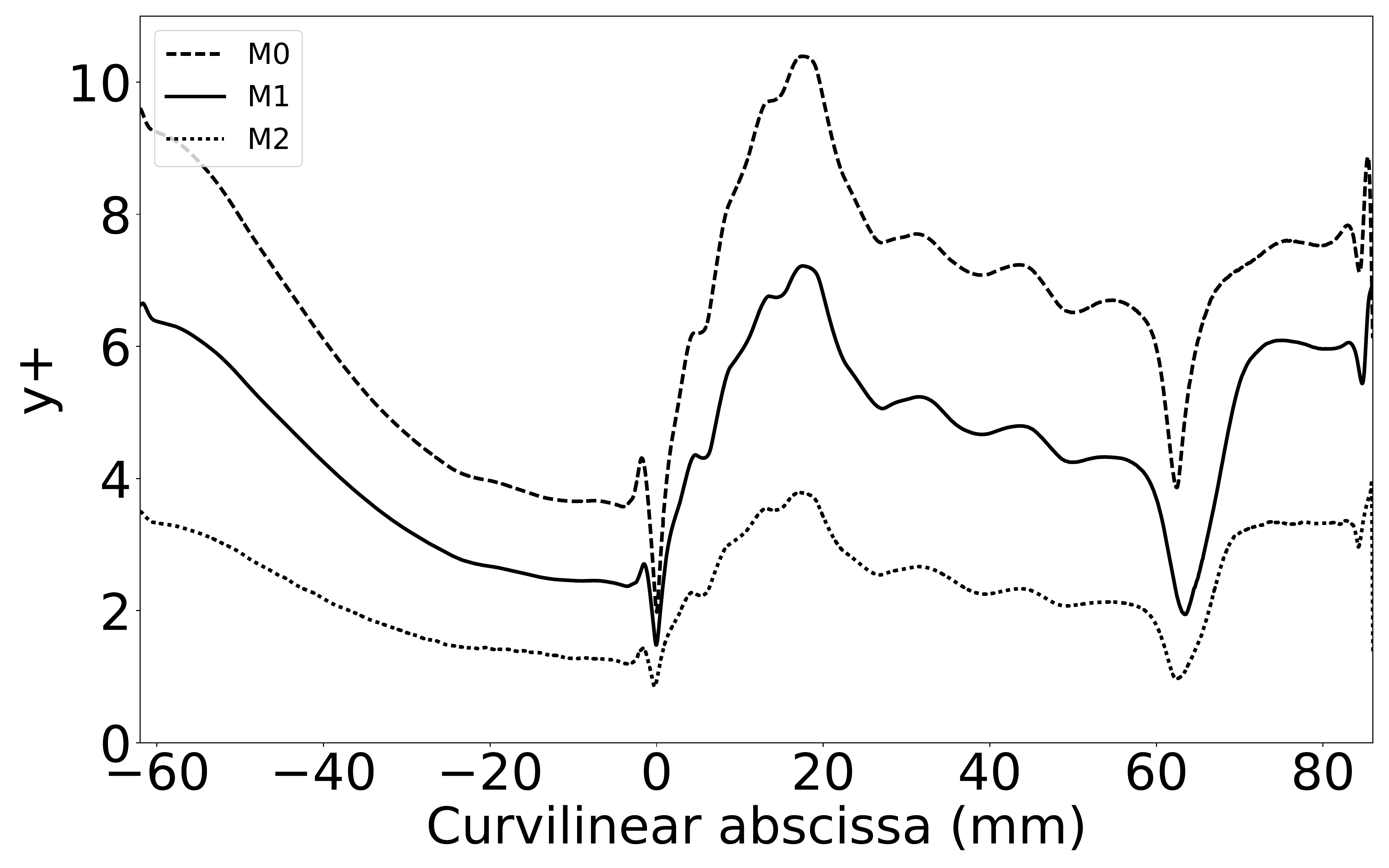}
\caption{Refinement over blade surface measured using non-dimensional $y^+$ parameter for MUR235 operating point ($Tu=6\%, \alpha = 0\degree$).}
\label{fig:M0M1M2_y+comp}
\end{figure*}

The resolution of the mesh and the LES quality must be guaranteed to be sufficient to capture the complex physics encountered. Indeed, the interaction between the free-stream turbulence and the boundary layer requires to carefully mesh the near-wall region. It is reasonable then to compare the profiles of heat transfer obtained using the mesh for this UQ study, from here on denoted as \textit{M0}, to two finer meshes \text{M1} and \textit{M2}, see ~\cref{fig:M0M1M2_hcomp}. The corresponding spatial distributions of ${y^+}$ are shown in~\cref{fig:M0M1M2_y+comp} for the three meshes.

The heat transfer coefficient is seen to be different on the pressure side for the finest mesh (M2). However, on the suction side the coarser mesh (M0) leads to approximately the same results as the finest mesh (M2). This suggests that the value of $y^+$ does not have a first order effect on the heat transfer coefficient for the meshes considered. The sensitivity to other effects such as turbulence intensity and angle of attack may thus be sought. Additionally, it can be noted that the shock wave on the suction side is located at approximately the same position for all meshes. This implies that the upstream boundary layer is similar in all cases although the heat transfer coefficient across the shock wave is affected by the mesh refinement.


\subsubsection{Uncertainty Quantification results}

This section presents the comparison between the different resampling methods on this complex case. In the following, an existing sample comprised of 16 simulations is used to generate a \textit{Sobol'} low-discrepancy sequence. As seen in~\cref{sec:doe}, the quality of \textit{Sobol'} sequence is similar to \textit{Halton}'s in low dimensional cases. Using this initial set of simulations, the sequence has been continued adding 4 points to give a total of 20 simulations. Then using the same initial sample, the previous set is compared to the use of the $\sigma$ method and the LOO-\textit{Sobol'} method. The LOO-$\sigma$ method gives similar results compared to LOO-\textit{Sobol'} method. It is not tested on this case. Quality results evaluated by LOO as described in~\cref{sec:error} are shown in~\cref{tab:ls89-q2}. 

\begin{table}[h]
\centering
\begin{tabular}{lcc}
\toprule
Method & Number of Simulations &$\hat{Q}_2$\\
\cmidrule{2-3}
\textit{Sobol'}&16 & 0.638\\
\textit{Sobol'}&20& 0.821\\
\textit{$\sigma$}&20& 0.688\\
LOO-\textit{Sobol'}&20& 0.856\\
\bottomrule
\end{tabular}
\caption{Estimated $Q_2$ function of the resampling method compared to an initial sample of 16 simulations.}
\label{tab:ls89-q2}
\end{table}

\begin{figure}[!h]
\centering
\includegraphics[width=0.9\linewidth]{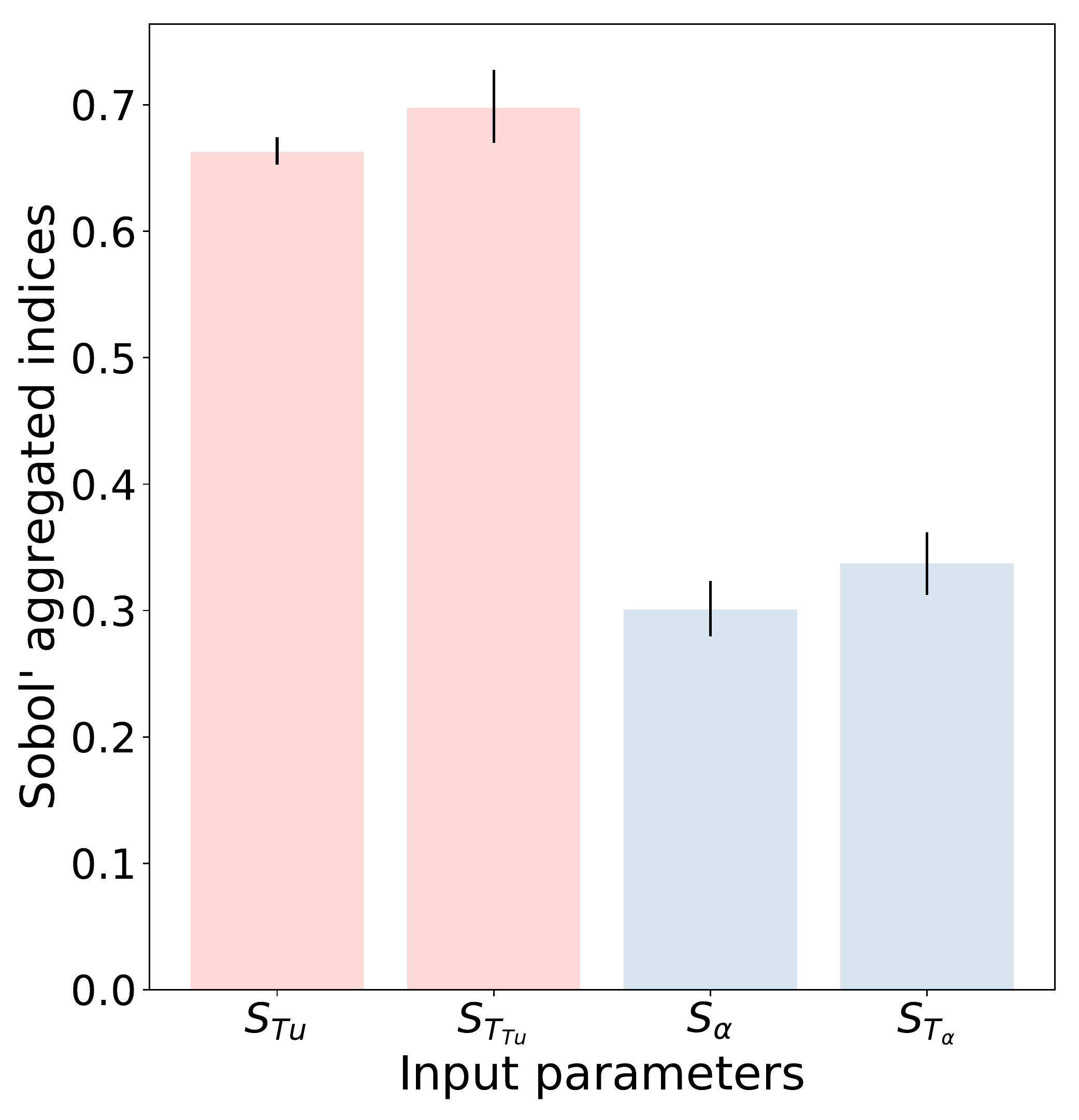}
\caption{Aggregated \textit{Sobol'} indices of the input parameters with their asymptotic confidence intervals.}
\label{fig:ls89-aggregated}
\end{figure}

As demonstrated in~\cref{sec:functions}, there is no guarantee that the quality of the model improves when using a refinement strategy other than continuing the low discrepancy sequence, given a low-dimensional case. The $\sigma$ method was only able to improve a little the quality of the initial design. This improvement was inferior to the simple continuation of the sequence. However, we observed an improved quality using the LOO-\textit{Sobol'} method. The importance factors' difference between the two input parameters make it feasible to improve further the quality of the model---see~\cref{fig:ls89-aggregated}.

\begin{figure*}[!h]
\centering
\subfloat[\textit{Sobol'} sequence]{
\includegraphics[width=0.47\linewidth,height=\textheight,keepaspectratio]{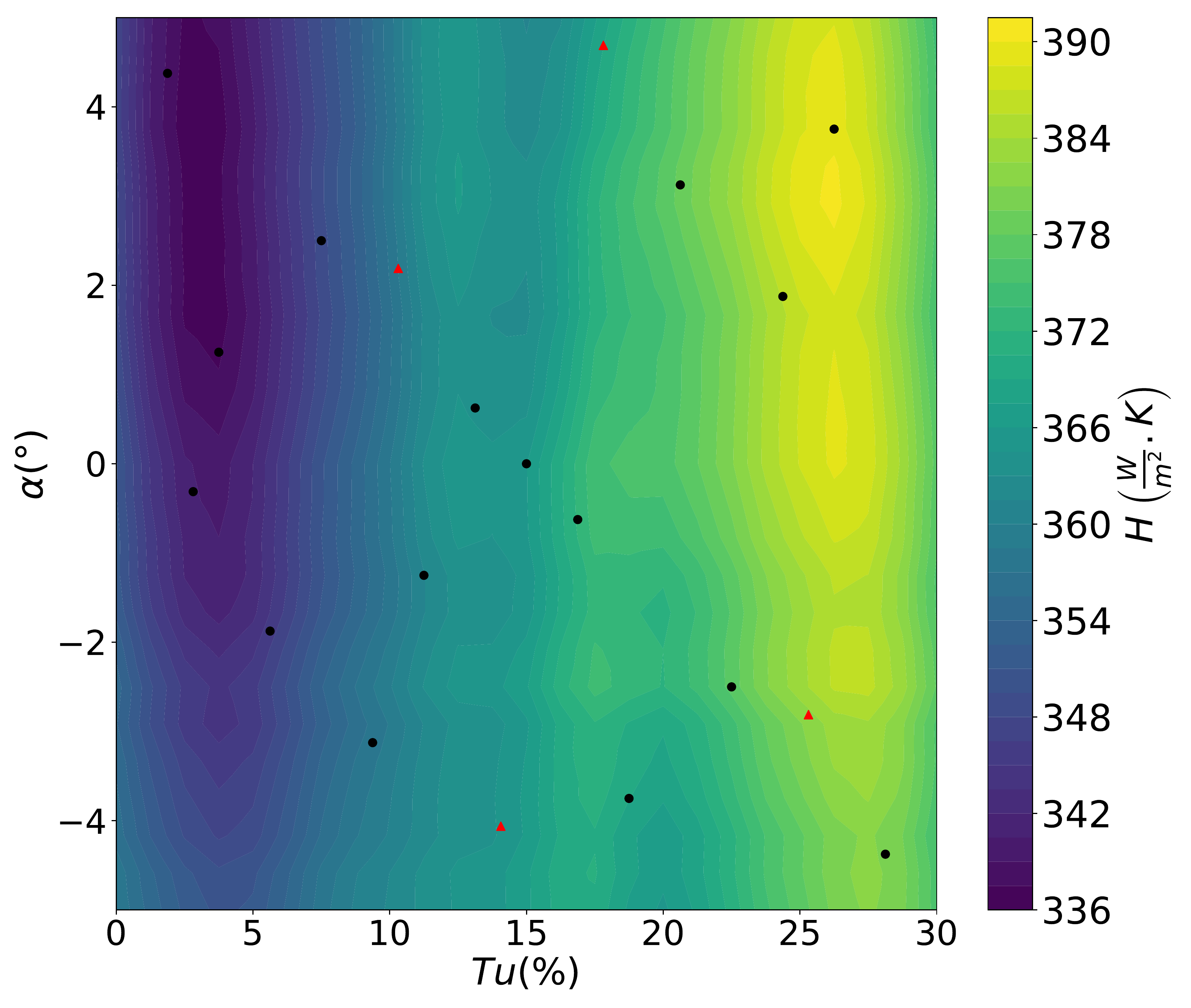}}
~
\subfloat[$\sigma$ method]{
\includegraphics[width=0.47\linewidth,height=\textheight,keepaspectratio]{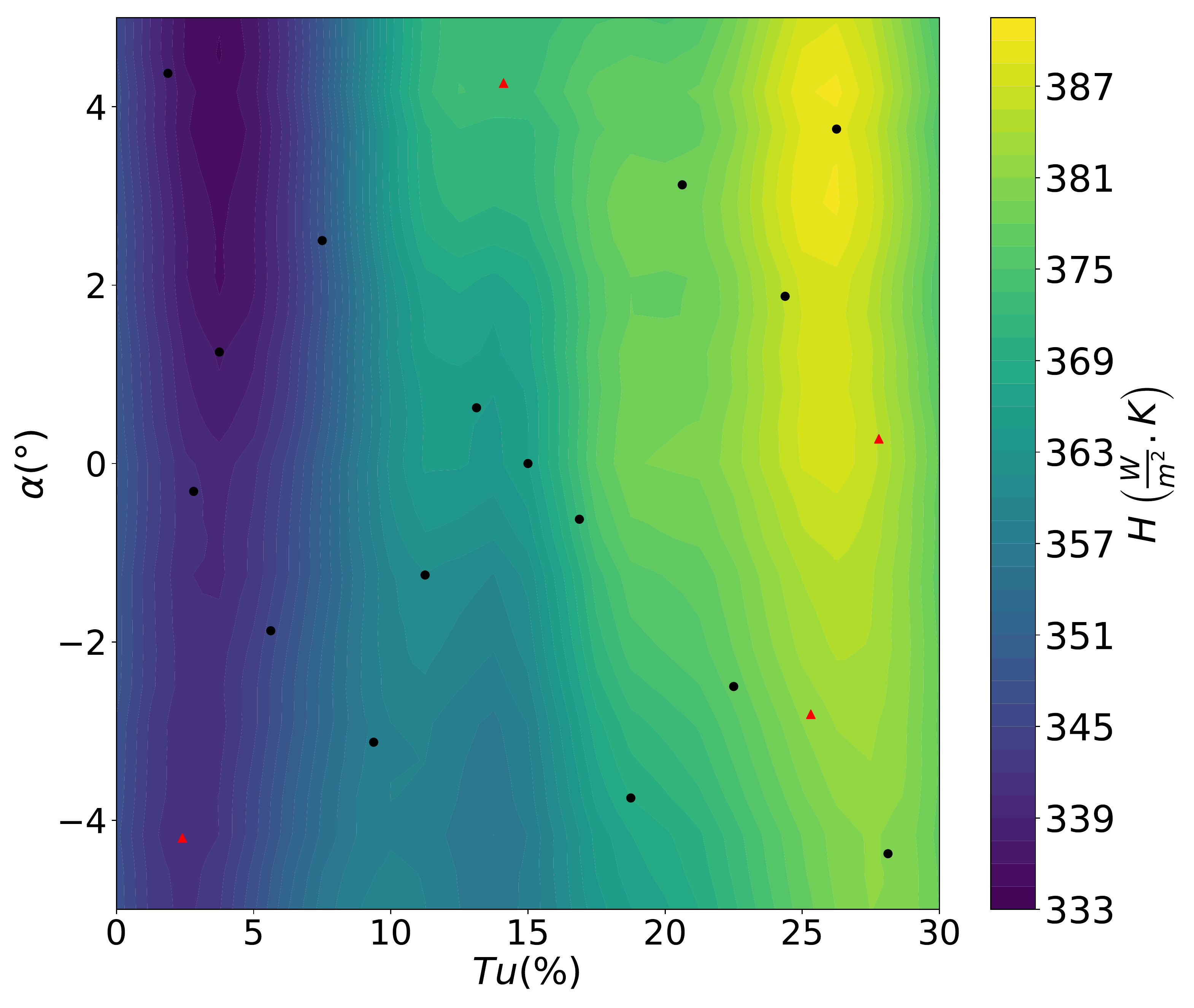}}

\subfloat[LOO-\textit{Sobol'} method]{
\includegraphics[width=0.47\linewidth,height=\textheight,keepaspectratio]{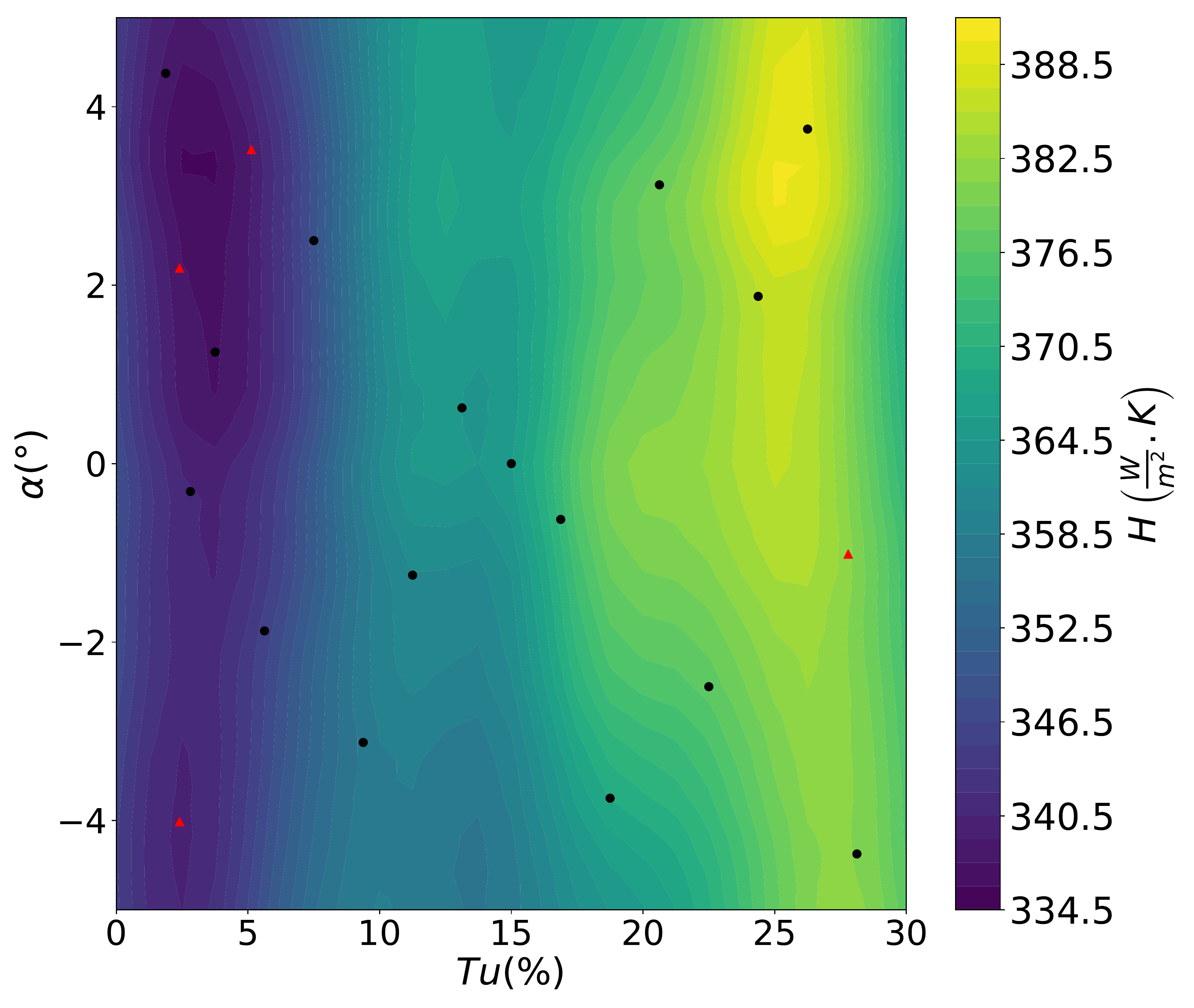}}
\caption{Heat Transfer coefficient response surface. DoE is initially composed of 16 simulations sampled with \textit{Sobol'} sequence. Dots represent the initial LES simulations and diamonds represent the resampled points.}
\label{fig:ls89-RS}
\end{figure*}

The response surfaces of the models are plotted in \cref{fig:ls89-RS}. The heat transfer coefficient has been integrated over the chord line to obtain this visualization. The first thing to notice is the correct distribution of sample points within the parameter space ensuring that most of the effects are captured. The predictions obtained using the models are then found to be in agreement with the observations made previously. The heat transfer coefficient increases with the turbulence intensity and is fairly stable regarding the angle of the incoming flow. The models are said to be additive with respect to the turbulence intensity. Contrary to the \textit{Sobol'} sequence, the LOO-\textit{Sobol'} method detected that the model was sensitive to low values of turbulence intensity. It is this physical information that helped improve the predictivity quality. In the following, the model constructed using the LOO-\textit{Sobol'} method is used.

\begin{figure*}[!h]
\centering
\includegraphics[width=0.9\linewidth]{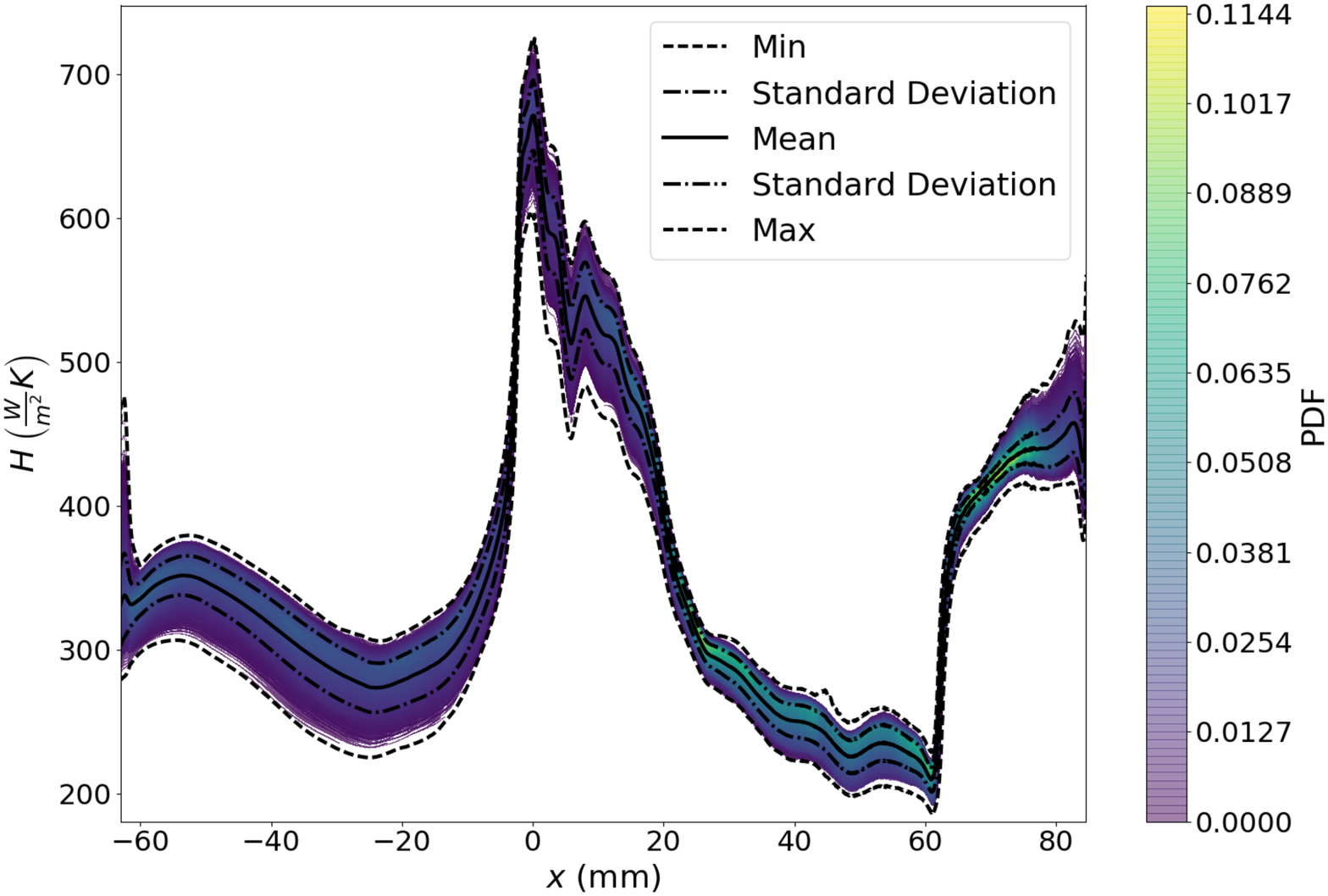}
\caption{Probability Density Function and moments of the heat transfer coefficient along the chord line of the blade.}
\label{fig:ls89-propagation}
\end{figure*}

Without making any assumption on the uncertainties, the Probability Density Functions (PDF) of the input parameters are both defined using uniform distributions over the parameter space
\begin{align}
Tu\sim \mathcal{U}(0,30\%) \quad \alpha \sim \mathcal{U}(-5,5\degree).
\end{align}
Using these PDFs, uncertainties are propagated by $\numprint{5000}$ predictions of the heat transfer coefficient along the blade. Then the QoI's PDF is reconstructed using a kernel smoothing procedure~\cite{wand1995,hastie2009}. \Cref{fig:ls89-propagation} reveals the expected concerning the propagation of such uncertainties to the heat transfer coefficient. As the two input distributions are uniform and the model is additive, the mean is centred between the extrema. From the experiments---see~\cref{fig:space-tu-alpha}---the envelope of the heat transfer coefficient is correctly captured except after the shock region. Indeed, from past experiences, capturing this region requires a value of $y^+\sim 1-2$~\cite{Segui2017b}.

\begin{figure*}[!h]
\centering
\includegraphics[width=0.9\linewidth]{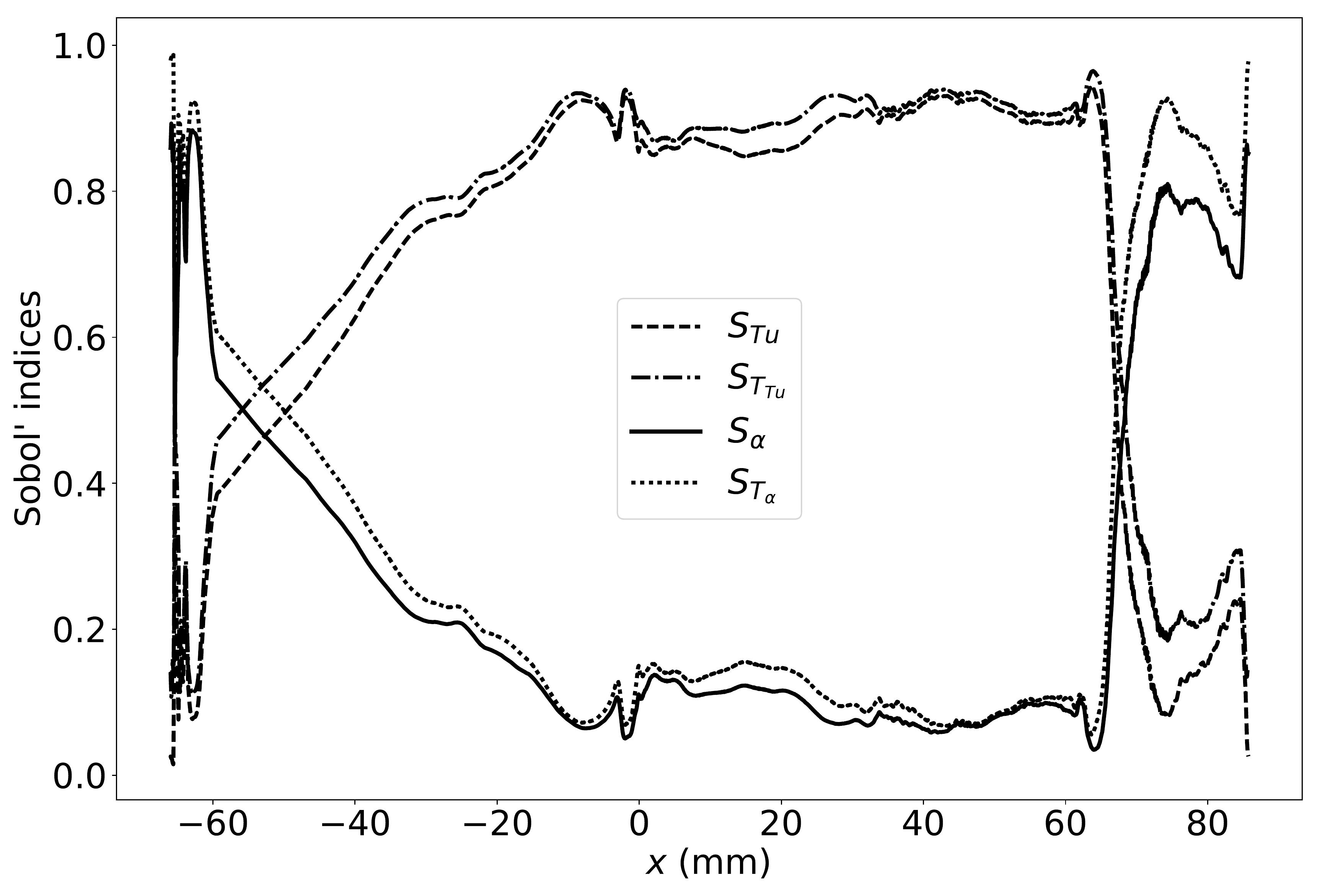}
\caption{First order and total order \textit{Sobol'} indices along the chord line.}
\label{fig:ls89-map}
\end{figure*}

Finally, the \textit{Sobol'} indices have been estimated using $\numprint{200000}$ predictions. As the response surface suggested, the heat transfer coefficient is mainly affected by the variation of the turbulence intensity. The spatial evolution of the indices in~\cref{fig:ls89-map}, shows a spatial dependency. On the pressure side, the inflow angle has a higher influence as its contribution rises to become the most important parameter at the trailing edge. On the suction side, the turbulence intensity contribution is stable until the shock region. Reaching the trailing edge, the angle contribution increases. Finally, aggregated indices are reported in~\cref{fig:ls89-aggregated}. These indices confirm that the turbulence intensity is the most important parameter compared to the inflow angle when studying the heat transfer coefficient and for the range of angle variations retained. The turbulence intensity contributes to 70\% of the total variance of the QoI whereas the inflow angle contributes to 30\%. This behaviour was expected as downstream the shock, the incoming level of turbulence has little impact. The computation of the second order indices are not presented here because their values are negligible in comparison to the first order indices. This is in agreement with the small differences observed between the first and total order indices. There are no joint effects between the two parameters.

\section{Conclusions}
\label{sec:ccl}

Two new methods have been introduced in this work for resampling the parameter space in order to improve the predictivity coefficient of a surrogate model: namely LOO-$\sigma$ and LOO-\textit{Sobol'} methods. These methods do not only take advantage of the capability of Gaussian Process models to infer a prediction variance, but they use information about the observed quality of the model. It was shown that an improvement of the quality of the model is guaranteed in high dimensional cases. Compared to a resampling method based on the predicted variance only, the proposed methods behaviour appears to be more stable and reliable. We also found that the ratio of the initial learning sample space over the total budget of function evaluation should remain greater than $\numprint{0.5}$. Which is to say that no more than half of the budget should be allocated to resampling the parameter space. In any case, the initial quality of the model should be reasonable when considering these techniques.

A first Uncertainty Quantification LES study of the \textit{LS89} is presented. The parameter space was comprised of the turbulence intensity and the inflow angle. In order to increase the quality of the surrogate model, the LOO-\textit{Sobol'} method was used to refine the parameter space. We showed that it performed better than continuing the sampling sequence. Apart from an analysis of the variance, the model was used to propagate uncertainties. This study reveals that although the turbulence intensity is the main factor impacting the heat transfer coefficient, there is spatial evolution of its contribution along the blade.

In terms of conclusions, by taking into account the physics in this process, the proposed methods will help build better models at lower costs. This will allow also Uncertainty Quantification of high-dimensional or expensive cases to be within reach.

\section*{Acknowledgments}
The financial support provided by all the CERFACS shareholders (AIRBUS Group, Cnes, EDF, Météo-France, ONERA, SAFRAN and TOTAL) is greatly appreciated and we thank them to enable the achievement of such research activities. Furthermore, the authors acknowledge GENCI [CCRT/CINES/IDRIS] for giving access to HPC resources under the allocation x20162a6074. These resources have permitted the preliminary studies of the LS89.

\bibliography{wileyNJD-AMA}

\end{document}